\documentclass[a4paper]{article}
\usepackage{pdfsync}
\usepackage[latin1]{inputenc}
\usepackage{graphicx}
\usepackage{amssymb,amsfonts,amsmath}
\usepackage{theorem}
\usepackage{hyperref}
\usepackage{graphicx}
\usepackage{multirow}
\usepackage{verbatim}

\usepackage{color}

\newtheorem{corollary}{Corollary}[section]

\newtheorem{proposition}{Proposition}[section]
{\theorembodyfont{\rmfamily}

\newtheorem{remark}{Remark}[section]
\newtheorem{example}{Example}[section]
}

\begin{document}

\title{\textbf{Relative velocities for radial motion in expanding Robertson-Walker spacetimes}}

\author{V. J. Bol\'os$^1$, D. Klein$^2$
\\
{\small $^1$ Dpto. Matem\'aticas para la Economía y la Empresa, Facultad de Econom\'{\i}a,}\\ {\small Universidad de Valencia. Avda. Tarongers s/n. 46022, Valencia,
Spain.}\\ {\small e-mail\textup{: \texttt{vicente.bolos@uv.es}}}\\
\\
{\small $^2$ Dept. Mathematics and Interdisciplinary Research Institute for the Sciences,}\\ {\small California State University, Northridge, USA}\\ {\small e-mail\textup{: \texttt{david.klein@csun.edu}}}}

\date{June 2011}

\maketitle

\begin{abstract}
The expansion of space, and other geometric properties of cosmological models, can be studied using geometrically defined notions of relative velocity.  In this paper, we consider test particles undergoing radial motion relative to comoving (geodesic) observers in  Robertson-Walker cosmologies, whose scale factors are increasing functions of cosmological time.   Analytical and numerical comparisons of the Fermi, kinematic, astrometric, and the spectroscopic relative velocities of test particles are given under general circumstances. Examples include recessional comoving test particles in the de Sitter universe, the radiation-dominated universe, and the matter-dominated universe. Three distinct coordinate charts, each with different notions of simultaneity, are employed in the calculations. It is shown that the astrometric relative velocity of a radially receding test particle cannot be superluminal in any expanding Robertson-Walker spacetime. However, necessary and sufficient conditions are given for the existence of superluminal Fermi speeds, and it is shown how the four concepts of relative velocity determine geometric properties of the spacetime.
\end{abstract}



\section{Introduction}

General relativity provides no \textit{a priori} definition of relative velocities of non local objects in curved spacetime.  Different coordinate systems and notions of relative velocity yield different results for the motion of distant test particles relative to a particular observer.  This ambiguity led to consideration of the need for a strict definition of \textquotedblleft radial
velocity\textquotedblright\ at the General Assembly of
the International Astronomical Union (IAU), held in 2000 (see
\cite{Soff03,Lind03}).

Thereafter, a series of papers \cite{Bolos02,Bolos05,Bolos07} appeared addressing the general question of relative velocities  and culminated in the introduction of four geometrically defined (but inequivalent) notions of relative velocity: Fermi, kinematic, astrometric, and the spectroscopic relative velocities. These four relative velocities each have physical justifications, and have been employed for studying properties of spacetimes \cite{KC10,Klein11}. Related work includes \cite{Carrera10b} for the study of observer-referred kinematics and dynamics,
\cite{Doran08} for a study of Schwarzschild black holes, and \cite{Lachieze06,Carrera10} for Doppler tracking.

The four definitions of relative velocities depend on two different notions of simultaneity: ``spacelike simultaneity'' (or ``Fermi simultaneity'' \cite{Ferm22}) as defined by Fermi coordinates of the observer, and ``lightlike simultaneity'' as defined by optical (or observational) coordinates of the observer \cite{Elli85}. The Fermi and kinematic relative velocities can be described in terms of the former, according to which events are simultaneous if they lie on the same space slice determined by Fermi coordinates. The kinematic relative velocity is found by first parallel transporting the 4-velocity $u'$ of the test particle at the spacetime point $q_{\mathrm{s}}$, along a radial spacelike geodesic (lying on a Fermi space slice) to a 4-velocity denoted by $\tau_{q_{\mathrm{s}}p}u'$ in the tangent space of the central observer at spacetime point $p$, whose 4-velocity is $u$. The kinematic relative velocity $v_{\mathrm{kin}}$ is then the unique vector orthogonal to $u$, in the tangent space of the observer, satisfying $\tau_{q_{\mathrm{s}}p}u'=\gamma (u+v_{\mathrm{kin}})$ for some scalar $\gamma$ (which is uniquely determined). For a test particle undergoing radial motion, the Fermi relative velocity, $v_{\mathrm{Fermi}}$ is the rate of change of proper distance of the test particle away from the central observer along the Fermi space slice, with respect to proper time of the observer.

The spectroscopic (or barycentric) and astrometric relative velocities can be found from spectroscopic and astronomical observations.  Mathematically, both rely on the notion of ``lightlike simultaneity'', according to which two events are simultaneous if they both lie past-pointing horismos (which is tangent to the backward light cone) at the spacetime point $p$ of the central observer. The spectroscopic relative velocity $v_{\mathrm{spec}}$ is calculated analogously to $v_{\mathrm{kin}}$, described in the preceding paragraph, except that the 4-velocity $u'$ of the test particle is parallel transported to the tangent space of the observer along a null geodesic lying on the past-pointing horismos of the observer, instead of along the Fermi space slice.  The astrometric relative velocity, $v_{\mathrm{ast}}$, of a test particle whose motion is purely radial is calculated analogously to $v_{\mathrm{Fermi}}$, as the rate of change of the affine distance, which corresponds to the \textit{observed} proper distance (through light signals at the time of observation) with respect to the proper time of the observer, as may be done via parallax measurements. We describe this more precisely in the sequel, and complete definitions for arbitrary (not necessarily radial) motion may be found in \cite{Bolos07}.

In \cite{Klein11} exact Fermi coordinates were found for expanding Robertson-Walker spacetimes and were shown to be global in the non inflationary case. Fermi coordinates were then used to calculate the (finite) diameter of the Fermi space slice, as a function of the observer's proper time, and  Fermi velocities of (receding) comoving test particles.   In this paper, we extend the results of \cite{Klein11} by calculating all four relative velocities for test particles undergoing arbitrary radial motion in general expanding Robertson-Walker spacetimes.  Examples include explicit expressions for the Fermi, kinematic, astrometric, and the spectroscopic relative velocities of comoving test particles in the de Sitter universe, the radiation-dominated universe, the matter-dominated universe, and more generally, cosmologies for which the scale factor, $a(t)=t^{\alpha}$ with $0<\alpha\leq1$ (see equation \eqref{eqds2} below).

We express the metric tensor in optical coordinates to calculate the two relative velocities associated with lightlike simultaneity.  Fermi coordinates are utilized to find functional relationships between all four relative velocities.  Necessary and sufficient conditions are given for the existence of superluminal Fermi relative velocities of test particles undergoing radial motion, and the astrometric relative velocity of a radially receding test particle is shown to be necessarily subluminal (the other two relative velocities are subluminal by their definitions).  In addition we show how the relative velocities determine the leading coefficients of the metric tensor in Fermi and optical coordinates (denoted as $g_{\tau\tau}$ and $\tilde{g}_{\tau\tau}$ below) and that pairs of them determine the scale factor, $a(t)$, of the Robertson-Walker spacetime.

This paper is organized as follows.  In Section \ref{sec:2}, we establish notation, define the four relative velocities for our circumstances, express the metric in Fermi and optical coordinate systems, derive general properties of the relative velocities and explain how they may be compared.  In Section \ref{comoving} we specialize to the case of test particles comoving with the Hubble flow and then apply the formulas in Section \ref{examples} to the Milne universe, the de Sitter universe, and to Robertson-Walker spacetimes for which the scale factor has the form $a(t)=t^{\alpha}$ with $0<\alpha<1$. This latter class of cosmologies includes the radiation-dominated and matter-dominated universes.  In Section \ref{sec:5}, we again consider general Robertson-Walker spacetimes and find functional relationships between the four relative velocities and the underlying geometry through the metric tensor.  Section \ref{conclusion} gives concluding remarks with a discussion of relative velocities and expansion of space.

\section{Relative velocities}
\label{sec:2}

The Robertson-Walker metric in curvature-normalized coordinates (or Robertson-Walker coordinates) is given by the line element
\begin{equation}
\label{eqds2}
\mathrm{d}s^2=-\mathrm{d}t^2+a^2(t)\left( \mathrm{d}\chi ^2 +S_k^2\left( \chi \right) \mathrm{d}\Omega ^2 \right) ,
\end{equation}
where $\mathrm{d}\Omega =\mathrm{d}\theta +\sin ^2\theta \mathrm{d}\varphi $, $a(t)$ is a positive and increasing scale factor, with $t>0$, and
\begin{equation*}
S_k\left( \chi \right) :=\left\{
\begin{array}{ll}
  \sin \left( \chi \right) & \textrm{if }k=1 \\
  \chi & \textrm{if }k=0 \\
  \sinh \left( \chi \right) & \textrm{if }k=-1.
\end{array}
\right.
\end{equation*}
There is a coordinate singularity in \eqref{eqds2} at $\chi =0$, but this will not affect the calculations that follow. Since our purpose is to study radial motion with respect to a central observer, it suffices to consider the $2$-dimensional Robertson-Walker metric given by
\begin{equation}\label{metric}
\mathrm{d}s^2=-\mathrm{d}t^2+a^2(t)\mathrm{d}\chi ^2,
\end{equation}
for which there is no singularity at $\chi=0$.

\subsection{Notation}
\label{notation}

We denote a central observer by $\beta (\tau ):=(\tau ,0)$, and a test particle by $\beta ' \left( \tau '\right) =\left( t\left( \tau '\right) , \chi \left( \tau '\right) \right) $. Both spacetime paths are parameterized by their proper times, and we will always assume that $\chi >0$. Our aim is to study the relative velocities of $\beta ' $ with respect to, and observed by, $\beta $.
The 4-velocity of $\beta $ is denoted by $U:=\frac{\partial }{\partial t}=(1,0)$, and the 4-velocity of $\beta ' $ is given by $U':=\dot{t}\frac{\partial }{\partial t}+\dot{\chi }\frac{\partial }{\partial \chi }=( \dot{t},\dot{\chi })$, where the overdot represents differentiation with respect to $\tau '$, the proper time of $\beta ' $. From $g(U',U')=-1$, we obtain
\begin{equation}
\label{dott}
\dot{t}=\sqrt{a^2(t)\dot{\chi }^2+1}.
\end{equation}

Vector fields will be represented by upper case letters, and vectors by lower case letters. Following this notation, the 4-velocity of $\beta $ at a fixed event $p=(\tau ,0)$ will be denoted by $u=(1,0)$. The Fermi, kinematic, spectroscopic, and astrometric relative velocity vector fields (to be defined below) are denoted respectively as $V_{\mathrm{kin}}$, $V_{\mathrm{Fermi}}$, $V_{\mathrm{spec}}$ and $V_{\mathrm{ast}}$.  They are vector fields defined on the spacetime path, $\beta $, i.e., the central observer.  Since all of these relative velocities are spacelike and orthogonal to $U$, they are each proportional to the unit vector field $\mathcal{S}:=\frac{1}{a(t)}\frac{\partial }{\partial \chi }$.\footnote{This should not to be confused with the relative position vector field $S$ used in \cite{Bolos07}; in fact, $\mathcal{S}$ is the normalized version of $S$.}

\begin{figure}[tbp]
\begin{center}
\includegraphics[width=0.35\textwidth]{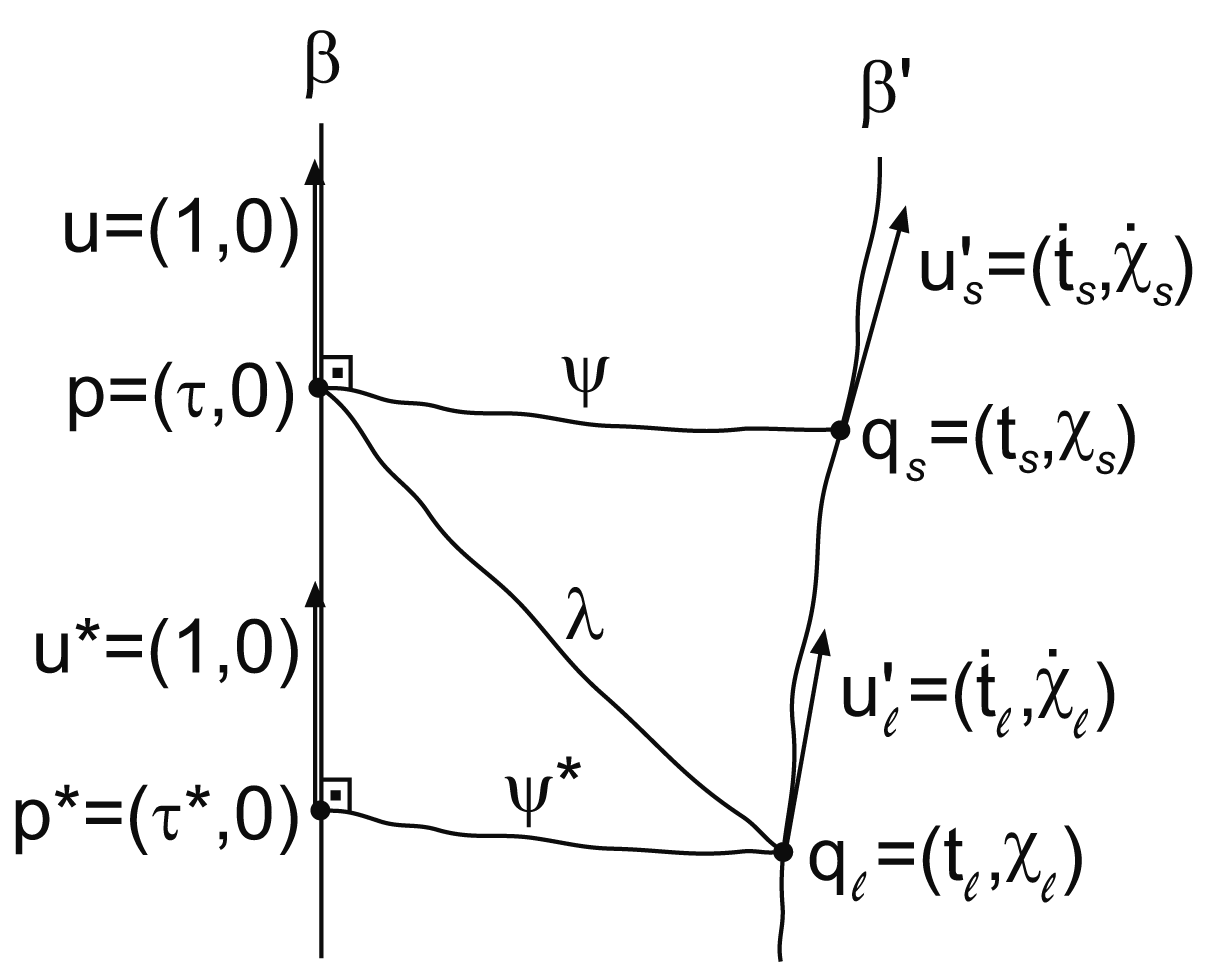}
\end{center}
\caption{Scheme of the elements involved in the study of the relative velocities of a test particle $\beta ' $ with respect to the central observer $\beta $. The curves $\psi $ and $\psi ^*$ are spacelike geodesics orthogonal to the 4-velocity of $\beta $, and $\lambda $ is a lightlike geodesic.}
\label{diagram}
\end{figure}

Comparisons of the four relative velocities for a given test particle are possible.  Direct comparisons may be made of the Fermi and kinematic relative velocities, because of the common dependence of these two notions of relative velocity on spacelike simultaneity.  Similarly, direct comparisons of the astrometric and spectroscopic relative velocities are also possible.  However, a comparison of all four relative velocities made at a particular instant by the central observer $\beta $ is possible only with data from two different spacetime events ($q_{\mathrm{s}}$ and $q_{\ell}$ in Figure \ref{diagram}) of the test particle.  Such a comparison, to which we refer as an \emph{instant comparison}, therefore lacks physical significance, unless the evolution of the test particle $\beta ' $ can be deduced from its 4-velocity at one spacetime point, e.g. for comoving or, more generally, geodesic test particles.

It is also possible to compare all four relative velocities at a fixed spacetime event $q_{\ell}$ of the test particle through observations from two different times of the central observer, identified as $\tau$ and $\tau^*$ in Figure \ref{diagram}. In the sequel, we refer to such a comparison of the relative velocities as a \emph{retarded comparison}.

In all that follows, we use the following notation for vectors at a given spacetime point $p=(\tau,0)$ and a spacetime point $p^*=(\tau^*,0)$ in the past of $p$ (see Figure \ref{diagram}): $v_{\mathrm{kin}}:=V_{\mathrm{kin}~p}$, $v_{\mathrm{Fermi}}:=V_{\mathrm{Fermi}~p}$, $v_{\mathrm{kin}}^*:=V_{\mathrm{kin}~p^*}$, $v_{\mathrm{Fermi}}^*:=V_{\mathrm{Fermi}~p^*}$, $v_{\mathrm{spec}}:=V_{\mathrm{spec}~p}$ and $v_{\mathrm{ast}}:=V_{\mathrm{ast}~p}$.
So, in an instant comparison we compare $v_{\mathrm{kin}}$, $v_{\mathrm{Fermi}}$, $v_{\mathrm{spec}}$, and $v_{\mathrm{ast}}$, while in a retarded comparison we compare $v_{\mathrm{kin}}^*$, $v_{\mathrm{Fermi}}^*$, $v_{\mathrm{spec}}$, and $v_{\mathrm{ast}}$.

\subsection{Spacelike simultaneity and Fermi coordinates}
\label{seckinf1}

Let $p=(\tau ,0)$ be an event of the central observer $\beta $ with $4$-velocity $u$. An event $q$ is \textit{spacelike} (or \textit{Fermi}) \textit{simultaneous} with $p$ if $g(\exp_p^{-1}q,u)=0$\footnote{The exponential map, $\exp_{p}v$, denotes the evaluation at affine parameter $1$ of the geodesic starting at the point $p$, with initial derivative $v$.}. The \textit{Fermi space slice} $\mathcal{M}_{\tau}$ (see \cite{Klein11}, it is called \textit{Landau submanifold} and denoted by $L_{p,u}$ in \cite{Bolos02}) is composed of all the events that are spacelike simultaneous with $p$.

More explicitly, the vector field
\begin{equation}
\label{eqX}
X:=-\sqrt{\left( \frac{a(\tau )}{a(t)}\right) ^2-1}\frac{\partial }{\partial t}+\frac{a(\tau )}{a^2(t)}\frac{\partial }{\partial \chi}
\end{equation}
is geodesic, spacelike, unit, and $X_p$ is orthogonal to the 4-velocity $u=(1,0)$ at $p$, i.e. $X_p$ is tangent to $\mathcal{M}_{\tau }$. Let $q_{\mathrm{s}}:=\left( t_{\mathrm{s}},\chi _{\mathrm{s}} \right) $ be the unique event of $\beta ' \cap \mathcal{M}_{\tau }$. Then there exists an integral curve of $X$ from $p$ to $q_{\mathrm{s}}$ (the geodesic $\psi $ in Figure \ref{diagram}), and so, using  \eqref{eqX} we can find a relationship between $\tau $, $t_{\mathrm{s}}$ and $\chi _{\mathrm{s}}$:
\begin{equation}
\label{eqt1} \int _{\tau }^{t_{\mathrm{s}}} \frac{a(\tau )}{a^2(\bar{\tau})}\frac{-1}{\sqrt{\left( \frac{a(\tau )}{a(\bar{\tau})}\right) ^2-1}}\,\textrm{d}\bar{\tau} =\int _0^{\chi _{\mathrm{s}}} \,\textrm{d}\bar{\chi}
\quad \Longrightarrow \quad
\int _{t_{\mathrm{s}}}^{\tau } \frac{a(\tau )}{a(\bar{\tau})}\frac{1}{\sqrt{a^2(\tau )-a^2(\bar{\tau})}}\,\textrm{d}\bar{\tau} =\chi _{\mathrm{s}}.
\end{equation}
From \eqref{eqt1} we obtain that $t_{\mathrm{s}}<\tau $. Moreover, since we only consider positive coordinate times, we have to impose $t_{\mathrm{s}}>0$ and then it is necessary that $0<\chi _{\mathrm{s}}<\chi _{\mathrm{smax}}$, where
\begin{equation}
\label{eqchiSmax}
\chi _{\mathrm{smax}}:=\int _{0}^{\tau } \frac{a(\tau )}{a(\bar{\tau })}\frac{1}{\sqrt{a^2(\tau )-a^2(\bar{\tau })}}\,\textrm{d}\bar{\tau }.
\end{equation}
On the other hand, with respect to time, it is necessary that $\tau >\tau _{\mathrm{smin}}$, where $\tau _{\mathrm{smin}}$ is defined implicitly by
\begin{equation}
\label{eqtSmin}
\int _{0}^{\tau _{\mathrm{smin}}} \frac{a(\tau )}{a(\bar{\tau })}\frac{1}{\sqrt{a^2(\tau )-a^2(\bar{\tau })}}\,\textrm{d}\bar{\tau }=\chi _{\mathrm{s}}.
\end{equation}
Note that $\chi _{\mathrm{smax}}(\tau _{\mathrm{smin}}(\chi _{\mathrm{s}}))=\chi _{\mathrm{s}}$ and $\tau _{\mathrm{smin}}(\chi _{\mathrm{smax}}(\tau ))=\tau $.

It will also be useful in what follows to consider Fermi coordinates.  In \cite{Klein11} the metric \eqref{eqds2} was expressed in Fermi coordinates with respect to the central observer $\beta $ (called Fermi observer in this context) for expanding Robertson-Walker spacetimes, and it was shown that Fermi coordinates are global when the spacetime is non inflationary. In two spacetime dimensions, the metric \eqref{metric} expressed in Fermi coordinates $(\tau, \rho)$ is given by
\begin{equation}\label{fermipolar}
ds^2=g_{\tau\tau} d\tau^2+d\rho^2,
\end{equation}
where $\rho$ is proper distance along a spacelike geodesic, orthogonal to the path $\beta $, i.e., the Fermi distance from $\beta $ to the corresponding event (see \cite{Bolos07}). A formula for $g_{\tau \tau }$ in terms of $(\tau ,\rho )$ is given in \cite[Theorem 2]{Klein11}.

The vector field $X$ given in \eqref{eqX} may be expressed in Fermi coordinates as $X=\partial/\partial\rho$. By design of Fermi coordinates, $\tau = t$ on the path $\beta (t)$ of the Fermi observer (where $\rho =0$), but the two time coordinates differ away from that path. The coordinate transformations between Fermi coordinates $(\tau, \rho)$ and curvature-normalized coordinates $(t,\chi)$ are given as integral expressions in \cite{Klein11}, but will not be needed here.

\begin{remark}\label{light}
Setting $ds^2=0$ in \eqref{fermipolar}, shows that the velocity  of a distant photon with spacetime coordinates $(\tau,\rho)$, relative to the Fermi observer $\beta $, is given by $|d\rho/d\tau|= \sqrt{-g_{\tau\tau}(\tau,\rho)}$.  Thus, the metric \eqref{fermipolar} may be understood as a natural generalization of the Minkowski metric when the speed of a photon depends on its spacetime coordinates, and may expressed as $ds^2=-c(\tau,\rho)^2 d\tau^2+d\rho^2$ where $c(\tau,\rho)$ is the Fermi speed of a photon at the spacetime point $(\tau,\rho)$ relative to the Fermi observer located at $(\tau,0)$.
\end{remark}

\subsection{Fermi and kinematic relative velocities}
\label{sec:2.2}

In this section, we find general expressions for the Fermi and kinematic velocities of a radially moving test particle, relative to the central observer $\beta $.  These formulas will be applied to more specialized circumstances in later sections.

Using the Fermi coordinates developed in the previous section, let $u'_{\mathrm{s}}=\left. \dot{\tau}\frac{\partial }{\partial \tau }\right| _{q_{\mathrm{s}}} +\left. \dot{\rho}\frac{\partial }{\partial \rho }\right| _{q_{\mathrm{s}}}=(\dot{\tau},\dot{\rho})$ be the 4-velocity of a radially moving test particle $\beta ' $ at a point $q_{\mathrm{s}}$ with Fermi coordinates $(\tau, \rho)$, where the overdot represents differentiation with respect to proper time of $\beta ' $, and $p=(\tau ,0)$ is the event of $\beta $ from which we measure the velocities, with 4-velocity $u=(1,0)$. Taking into account \cite[Proposition 3]{Bolos07}, the Fermi relative velocity of $u'_{\mathrm{s}}$ with respect to $u$ is given by
\begin{equation}\label{fermidefinition2}
v_{\mathrm{Fermi}}=\frac{d\rho}{d\tau}\mathcal{S}_p=\frac{\dot{\rho}}{\dot{\tau}}\mathcal{S}_p,
\end{equation}
where $\mathcal{S}_p= \partial/\partial\rho|_p$. The requirement that $g(u'_{\mathrm{s}},u'_{\mathrm{s}})=-1$ forces $\|v_{\mathrm{Fermi}}\|<\sqrt{-g_{\tau\tau}(\tau, \rho)}$, which, by Remark \ref{light}, has the physical interpretation that the right hand side of the inequality is the speed, $|d\rho/d\tau|$, of a distant photon with spacetime coordinates $(\tau,\rho)$, relative to the Fermi observer $\beta $.  It is an upper bound and limiting value for the Fermi speed of a massive particle. We therefore have:

\begin{proposition}\label{superluminal}
In an expanding Robertson-Walker spacetime, the Fermi relative velocity of a radially moving test particle at position $(\tau, \rho)$ satisfies $\|v_{\mathrm{Fermi}}\|<\sqrt{-g_{\tau\tau}(\tau, \rho)}$ and can therefore exceed the central observer's local speed of light ($c=1$) within a Fermi coordinate chart if and only if $-g_{\tau\tau}(\tau, \rho) >1$.
\end{proposition}
The following examples illustrate the proposition.
\begin{example}
\begin{enumerate}
\item[a)] For Milne spacetime, $-g_{\tau\tau}(\tau, \rho)\equiv 1$, the Fermi chart is global, and thus all Fermi relative speeds are subluminal.
\item[b)] For the de Sitter universe, $-g_{\tau\tau}(\tau, \rho)=\cos^2(H_0\rho)$, with $H_0\rho <\pi/2$ (see \cite{CM,KC3,Klein11}) where $H_0$ is the Hubble constant. The Fermi chart is valid up to the cosmological horizon of this spacetime. Thus, all Fermi relative velocities are less than the local speed of light.
\item[c)] For the radiation-dominated universe, i.e., for the case that $a(t)=\sqrt{t}$ in \eqref{eqds2}, the Fermi chart is global and
\begin{equation}\label{radgtt}
-g_{\tau\tau}(\tau, \rho)=\frac{1}{\sigma}\left(1+\sqrt{\sigma-1}\,\sec^{-1}\sqrt{\sigma}\right)^2,
\end{equation}
where $\sigma\geq1$ depends on $\rho$ and $\tau$. It may be shown that for any $\tau>0$, the least upper bound of $\rho$ is $\frac{\pi}{2}\tau$ and that $\sqrt{-g_{\tau\tau}(\tau, \rho)}\rightarrow \frac{\pi}{2}$ asymptotically as  $\rho\rightarrow\frac{\pi}{2}\tau$ \cite{Klein11}. Thus, Fermi relative speeds can exceed the speed of light in this spacetime, but are bounded above by $\frac{\pi}{2}$.
\end{enumerate}
\end{example}

An alternative expression for the Fermi relative velocity follows by expressing the 4-velocity of the test particle as
\begin{equation}\label{4vel}
u'_{\mathrm{s}}= (\dot{\tau},\dot{\rho})= \frac{\mathcal{E}}{\sqrt{-g_{\tau\tau}(\tau,\rho)}}\left. \frac{\partial}{\partial\tau}\right| _{q_{\mathrm{s}}}\pm \sqrt{\mathcal{E}^2-1}\left. \frac{\partial}{\partial\rho}\right| _{q_{\mathrm{s}}},
\end{equation}
where $\pm$ indicates the sign of $\dot{\rho}$, and $\mathcal{E}:=\sqrt{-g_{\tau\tau}(\tau,\rho)}\,\dot{\tau}$ is the energy per unit mass of the test particle as measured by an observer with fixed Fermi spatial coordinate $\rho$, i.e., by an observer with 4-velocity
\begin{equation}\label{u0}
u_0= \frac{1}{\sqrt{-g_{\tau\tau}(\tau,\rho)}}\left. \frac{\partial}{\partial\tau}\right| _{q_{\mathrm{s}}},
\end{equation}
so that $-g(u_0,u'_{\mathrm{s}})=\mathcal{E}$. Note that a geodesic path for the test particle is determined by the value of $\mathcal{E}$ at a single spacetime point.  With this notation, we can write
\begin{equation}
\label{eqmodvfermigen}
\|v_{\mathrm{Fermi}}\|=\sqrt{-g_{\tau\tau}(\tau,\rho)}\,\frac{\sqrt{\mathcal{E}^2-1}}{\mathcal{E}}.
\end{equation}
Let $\tau _{q_{\mathrm{s}} p}$ represent the parallel transport from $q_{\mathrm{s}}$ to $p$ by the unique geodesic joining $q_{\mathrm{s}}$ and $p$, i.e. $\psi $ in Figure \ref{diagram}. Using the relations $g\left( \tau _{q_{\mathrm{s}} p}u'_{\mathrm{s}},\partial/\partial\rho|_p\right) =g\left( u'_{\mathrm{s}},\partial/\partial\rho|_{q_{\mathrm{s}}}\right) $ and $g\left( \tau _{q_{\mathrm{s}} p}u'_{\mathrm{s}},\tau _{q_{\mathrm{s}} p}u'_{\mathrm{s}}\right) =-1$, , we can find $\tau _{q_{\mathrm{s}} p}u'_{\mathrm{s}}$, and then, by \cite[Equation (4)]{Bolos07}, obtain the kinematic relative velocity of $u'_{\mathrm{s}}$ with respect to $u$ as
\begin{equation} \label{vkinfermi}
v_{\mathrm{kin}}=\frac{1}{\sqrt{-g_{\tau\tau}(\tau,\rho)}}\frac{d\rho}{d\tau}\mathcal{S}_p.
\end{equation}
Thus, taking into account \eqref{fermidefinition2} and \eqref{eqmodvfermigen} we have
\begin{equation}\label{vkinE}
\|v_{\mathrm{kin}}\|=\frac{\sqrt{\mathcal{E}^2-1}}{\mathcal{E}}.
\end{equation}

Returning to the curvature-normalized coordinates, and exploiting the symmetry of this spacetime through the killing field $\partial/\partial\chi$, the kinematic relative velocity may be expressed explicitly in terms of $\dot{\chi}_{\mathrm{s}}$.  From $g\left( \tau _{q_{\mathrm{s}} p}u'_{\mathrm{s}},X_p\right) =g\left( u'_{\mathrm{s}},X_{q_{\mathrm{s}}}\right) $ and $g(\tau _{q_{\mathrm{s}} p}u'_{\mathrm{s}},\tau _{q_{\mathrm{s}} p}u'_{\mathrm{s}}) =-1$ we can obtain $\tau _{q_{\mathrm{s}} p}u'_{\mathrm{s}}$, and then find
\begin{equation}
\label{vkingeneral}
v_{\mathrm{kin}} = \frac{\dot{t}_{\mathrm{s}}\sqrt{\frac{a^2(\tau )}{a^2(t_{\mathrm{s}})}-1}+a(\tau )\dot{\chi }_{\mathrm{s}}}{\sqrt{\left( \dot{t}_{\mathrm{s}}\sqrt{\frac{a^2(\tau )}{a^2(t_{\mathrm{s}})}-1}+a(\tau )\dot{\chi }_{\mathrm{s}}\right) ^2+1}}\mathcal{S}_p ,
\end{equation}
where $\dot{t}_{\mathrm{s}}=\sqrt{a^2(t_{\mathrm{s}})\dot{\chi }_{\mathrm{s}}^2+1}$ by \eqref{dott}.

Also, from \eqref{fermidefinition2}, the Fermi relative velocity of $u'_{\mathrm{s}}$ with respect to $u$ is given by
\begin{equation}
\label{vfermigeneral}
v_{\mathrm{Fermi}} =\frac{\dot{\rho }}{\dot{\tau }}\mathcal{S}_p=\frac{\frac{\partial \rho }{\partial t_{\mathrm{s}}}\dot{t}_{\mathrm{s}}+\frac{\partial \rho }{\partial \chi _{\mathrm{s}}}\dot{\chi }_{\mathrm{s}}}{\frac{\partial \tau }{\partial t_{\mathrm{s}}}\dot{t}_{\mathrm{s}}+\frac{\partial \tau }{\partial \chi _{\mathrm{s}}}\dot{\chi }_{\mathrm{s}}}\mathcal{S}_p,
\end{equation}
where the function $\tau (t_{\mathrm{s}},\chi _{\mathrm{s}})$ is defined implicitly by \eqref{eqt1}, and
\begin{equation}
\label{eqrho}
\rho (t_{\mathrm{s}},\chi _{\mathrm{s}})=d_u^{\mathrm{Fermi}}(p,q_{\mathrm{s}})=\int _{t_{\mathrm{s}}}^{\tau (t_{\mathrm{s}},\chi _{\mathrm{s}})}\frac{a(\bar{\tau })}{\sqrt{a^2(\tau (t_{\mathrm{s}},\chi _{\mathrm{s}}))-a^2(\bar{\tau })}}\,\textrm{d}\bar{\tau },
\end{equation}
taking into account \eqref{eqX} and \cite[Proposition 2]{Bolos07}.

\subsection{Lightlike simultaneity and optical coordinates}
\label{secspecast1}

Let $p=(\tau ,0)$ be an event of the central observer $\beta $. An event is \textit{lightlike simultaneous} with $p$ if it lies on the past-pointing horismos $E^-_p$ (which is tangent to the backward light cone at the spacetime point $p$).

The vector field
\begin{equation}
\label{eqY}
Y:=-\frac{a(\tau )}{a(t)}\frac{\partial }{\partial t}+\frac{a(\tau )}{a^2(t)}\frac{\partial }{\partial \chi}
\end{equation}
is geodesic, lightlike, and the integral curve $\lambda$ such that $\lambda (0)=p$ is a past-pointing null geodesic, affinely parameterized that satisfies the hypotheses of \cite[Proposition 6]{Bolos07}. Let $q_{\ell}:=\left( t_{\ell},\chi _{\ell}\right) $ be the unique event of $\beta ' \cap E^-_p$. Then $\lambda $ is the unique geodesic from $p$ to $q_{\ell}$, and so, using \eqref{eqY} we can find a relationship between $\tau$, $t_{\ell}$ and $\chi _{\ell}$:
\begin{equation}
\label{eqt1b}
\int _{\tau }^{t_{\ell}} \frac{-1}{a(\bar{\tau })}\,\textrm{d}\bar{\tau }=\int _0^{\chi _{\ell}} \,\textrm{d}\bar{\chi }
\quad \Longrightarrow \quad
\int _{t_{\ell}}^{\tau } \frac{1}{a(\bar{\tau })}\,\textrm{d}\bar{\tau }=\chi _{\ell}.
\end{equation}
From \eqref{eqt1b} it follows that $t_{\ell}<\tau $. Moreover, since we consider only positive coordinate times, $t_{\ell}>0$, and then it is necessary that $0<\chi _{\ell}<\chi _{\ell \mathrm{max}}(\tau )$, where
\begin{equation}
\label{eqchiLmax}
\chi _{\ell \mathrm{max}}(\tau ):=\int _{0}^{\tau } \frac{1}{a(\bar{\tau })}\,\textrm{d}\bar{\tau }
\end{equation}
is the particle horizon for the observer $\beta $ at $p$. On the other hand, with respect to time, it is necessary that $\tau >\tau _{\ell \mathrm{min}}(\chi _{\ell})$, where $\tau _{\ell \mathrm{min}}(\chi _{\ell})$ is defined implicitly by
\begin{equation}
\label{eqtLmin}
\int _{0}^{\tau _{\ell \mathrm{min}}(\chi _{\ell})} \frac{1}{a(\bar{\tau })}\,\textrm{d}\bar{\tau }=\chi _{\ell}.
\end{equation}
Observe that $\chi _{\ell \mathrm{max}}(\tau _{\ell \mathrm{min}}(\chi _{\ell}))=\chi _{\ell}$ and $\tau _{\ell \mathrm{min}}(\chi _{\ell \mathrm{max}}(\tau ))=\tau $, so that each function is the inverse of the other.

In the framework of lightlike simultaneity, it will be convenient to use optical (or observational) coordinates with respect to the observer $\beta $. Referring to Figure \ref{diagram}, we denote the optical coordinates of the point $q_{\ell}=(t_{\ell},\chi _{\ell})$ by $\left( \tau ,\delta \right) $, where the affine distance $\delta$ from $p$ to $q_{\ell }$ is defined as the norm of the projection of $\exp_p^{-1}(q_{\ell })$ onto the orthogonal complement $u^{\bot}$ of $u$ (see \cite{Bolos07}). From \eqref{eqt1b}, $\tau (t_{\ell},\chi _{\ell})$ is determined implicitly, and differentiation gives
\begin{equation}
\label{partialtaul}
\frac{\partial \tau }{\partial t_{\ell}}=\frac{a(\tau )}{a(t_{\ell})},\qquad \qquad
\frac{\partial \tau }{\partial \chi _{\ell}}=a(\tau ).
\end{equation}
It follows from \eqref{eqY} and \cite[Proposition 6]{Bolos07} that
\begin{equation}
\label{delta}
\delta=\delta (t_{\ell},\chi _{\ell})=\int _{t_{\ell}}^{\tau (t_{\ell},\chi _{\ell})} \frac{a(\bar{\tau })}{a\left( \tau (t_{\ell},\chi _{\ell})\right) }\,\textrm{d}\bar{\tau }.
\end{equation}
Differentiating \eqref{delta} and using \eqref{partialtaul}, gives
\begin{equation}
\label{partialdeltal}
\frac{\partial \delta }{\partial t_{\ell}}=\frac{a(\tau )}{a(t_{\ell})}-\frac{a(t_{\ell})}{a(\tau )}-\delta \frac{\dot{a}(\tau )}{a(t_{\ell})},\qquad \qquad
\frac{\partial \delta }{\partial \chi _{\ell}}=a(\tau )-\delta \dot{a}(\tau ),
\end{equation}
where $\dot{a}(t)$ is the derivative of $a(t)$. Now using \eqref{partialtaul} and \eqref{partialdeltal}, we may express the Robertson-Walker metric in optical coordinates (with respect to $\beta $) in the form
\begin{equation}\label{opticalmetric}
ds^2=\tilde{g}_{\tau\tau}d\tau ^2+2d\tau d\delta\equiv -2\left( 1-\frac{\dot{a}(\tau)}{a(\tau)}\delta -\frac{1}{2}\frac{a^2\left( t_{\ell }\right)}{a^2(\tau )}\right)d\tau ^2+2d\tau d\delta ,
\end{equation}
where $t_{\ell}(\tau ,\delta )$ is given implicitly by \eqref{delta}.

\subsection{Astrometric and spectroscopic relative velocities}

Analogous to Section \ref{sec:2.2}, we find here general expressions for the astrometric and spectroscopic velocities of a radially moving test particle, relative to the central observer $\beta $.  The formulas will be used in later sections.

Using the optical coordinates developed in the previous section, let $u'_{\ell}=\dot{\tau }\frac{\partial }{\partial \tau }\vert _{q_{\ell }}+\dot{\delta }\frac{\partial}{\partial \delta }\vert _{q_{\ell }}=(\dot{\tau },\dot{\delta})$ be the 4-velocity of a radially moving test particle $\beta ' $ at a point $q_{\ell}$ with optical coordinates $(\tau, \rho)$ in the past-pointing horismos $E^-_p$, where the overdot represents differentiation with respect to proper time of $\beta ' $, and $p=(\tau ,0)$ is the event of $\beta $ from which we measure the velocities. From \cite[Proposition 7]{Bolos07}, \eqref{opticalmetric}, and taking into account that $g\left( u'_{\ell},u'_{\ell}\right) =-1$, the astrometric relative velocity of $u'_{\ell}$ with respect to $u$ is given by
\begin{equation}\label{vastoptical}
v_{\mathrm{ast}}=\frac{d\delta }{d\tau }\mathcal{S}_p=\frac{\dot{\delta }}{\dot{\tau }}\mathcal{S}_p = \frac{1}{2}\left(-\tilde{g}_{\tau\tau}-\frac{1}{\dot{\tau}^2}\right)\mathcal{S}_p.
\end{equation}
There is no upper bound for $\| v_{\mathrm{ast}}\|$ in the case of a radially approaching test particle (i.e. for the case $d\delta/d\tau<0$) because $\dot{\tau}$ can be chosen to be arbitrarily close to zero in \eqref{vastoptical}. However, a sharp upper bound for the case of a radially receding test particle is given by $\| v_{\mathrm{ast}}\| < -\tilde{g}_{\tau\tau}/2$, where the right side of this inequality is the relative speed, $d\delta/d\tau$, of a distant radially receding photon.  Since it follows from \eqref{opticalmetric} that $-\tilde{g}_{\tau\tau}<2$ when $\dot{a}(\tau)\geq0$, we have the following general result.

\begin{proposition}\label{subluminal}
In any expanding Robertson-Walker spacetime, the astrometric relative velocity of a radially receding test particle is always less than the central observer's local speed of light ($c=1$).
\end{proposition}

Using \eqref{opticalmetric}, it is easy to show that $\delta$ is an affine parameter for the null geodesic segment, $\lambda(\delta)=(\tau,\delta_0 - \delta)$ for a given $\delta _0>0$, where $\tau$ is fixed and $0\leq\delta\leq\delta_0$ (see Figure \ref{diagram}). Moreover, the vector field \eqref{eqY} is given by $Y=\partial/\partial\delta $ in optical coordinates. Observe that by design of optical coordinates, $\dot{\tau}=0$ along $\lambda(\delta)$.

Let $\nu $, $\nu '$ be the frequencies observed by $u$, $u'_{\ell }$, respectively, of a photon emitted from the spacetime point $q_{\ell }$. Then, the redshift of the test particle with radial motion relative to the central observer at $p$ is determined by the frequency ratio
\begin{equation}\label{doppleroptic}
\frac{\nu '}{\nu }=\frac{g(\textrm{P}_{q_{\ell }},u'_{\ell })}{g(\textrm{P}_p,u)},
\end{equation}
where $\textrm{P}=-\partial /\partial \delta $ is the 4-momentum tangent vector field of the emitted photon.
Using \eqref{opticalmetric} yields
\begin{equation}
\label{doppleroptic2}
\frac{\nu '}{\nu }=\dot{\tau},
\end{equation}
and thus,
\begin{equation}\label{vspecinoptical}
v_{\mathrm{spec}}=\frac{\dot{\tau}^2-1}{\dot{\tau}^2+1}\mathcal{S}_p.
\end{equation}
We may also use \eqref{doppleroptic2} along with \eqref{opticalmetric} in \eqref{vastoptical} to write an alternative expression for the astrometric relative velocity,
\begin{equation}\label{astrogeneral}
v_{\mathrm{ast}}=\left( 1-\frac{\dot{a}(\tau )}{a(\tau )}\delta -\frac{1}{2}\left( \frac{a^2(t_{\ell})}{a^2(\tau )}+ \left( \frac{\nu '}{\nu }\right) ^{-2}\right) \right) \mathcal{S}_p.
\end{equation}

In order to find expressions for the astrometric and spectroscopic relative velocities in terms of curvature-normalized coordinates, we make use of \eqref{partialtaul} to obtain
\begin{equation}\label{taudot}
\frac{\nu '}{\nu }=\dot{\tau}=\frac{\partial \tau }{\partial t_{\ell}}\dot{t}_{\ell}+\frac{\partial \tau }{\partial \chi _{\ell}}\dot{\chi }_{\ell} = a(\tau )\left( \sqrt{\dot{\chi }_{\ell}^2+a^{-2}(t_{\ell})}+\dot{\chi }_{\ell}\right) ,
\end{equation}
taking into account that $\dot{t}_{\ell}=\sqrt{a^2(t_{\ell})\dot{\chi }_{\ell}^2+1}$ by \eqref{dott}. Combining this last expression with \eqref{vspecinoptical} yields
\begin{equation}
\label{vspecgeneral2}
v_{\mathrm{spec}} =\frac{a^2(\tau )\left( \sqrt{\dot{\chi }_{\ell}^2+a^{-2}(t_{\ell})}+\dot{\chi }_{\ell}\right) ^2-1}{a^2(\tau )\left( \sqrt{\dot{\chi }_{\ell}^2+a^{-2}(t_{\ell})}+\dot{\chi }_{\ell}\right) ^2+1}\mathcal{S}_p.
\end{equation}
Similarly, combining \eqref{taudot} with \eqref{vastoptical} gives
\begin{equation}
\label{vastgeneral3}
v_{\mathrm{ast}}=\left(1- \frac{\dot{a}(\tau )}{a(\tau )}\delta-\frac{a^2(t_{\ell})}{2a^2(\tau )}
\left[1+\left( \sqrt{a^2(t_{\ell})\dot{\chi }_{\ell}^2+1}+a(t_{\ell})\dot{\chi }_{\ell}\right) ^{-2}\right]\right) \mathcal{S}_p.
\end{equation}

\begin{remark}
Expression \eqref{vastgeneral3} can be derived beginning with an expression analogous to \eqref{vfermigeneral} and using \eqref{partialtaul} and \eqref{partialdeltal}.
\end{remark}

\begin{remark}
\label{remretarded}
For ``retarded comparisons'', the kinematic and Fermi relative velocities of $u'_{\ell }$ must be calculated relative to $u^*=\left. \frac{\partial }{\partial t}\right| _{p^*}=(1,0)$, i.e., the 4-velocity of $\beta $ at $p^*=(\tau ^*,0)$. From \eqref{eqt1}, $\tau ^*=\tau^*\left( t_{\ell }(\tau ,\chi _{\ell }),\chi _{\ell }\right) $ is defined implicitly from
\begin{equation}
\label{eqtau*}
\int _{t_{\ell }(\tau ,\chi _{\ell })}^{\tau ^*} \frac{a(\tau ^*)}{a(\bar{\tau})}\frac{1}{\sqrt{a^2(\tau ^*)-a^2(\bar{\tau})}}\,\textrm{d}\bar{\tau} =\chi _{\ell },
\end{equation}
where $t_{\ell }(\tau ,\chi _{\ell })$ is given implicitly by \eqref{eqt1b}.
\end{remark}

\section{Comoving test particles}
\label{comoving}

In this section we apply the general results obtained in Section \ref{sec:2} to the case of test particles that are comoving with the Hubble flow.
A comoving test particle parameterized by its proper time is given by $\beta ' (\tau ')=(\tau ',\chi) $, where $\chi >0$ is constant, so that its 4-velocity $U'=(1,0)$ and therefore $u'_{\mathrm{s}}=(1,0)$ and $u'_{\ell}=(1,0)$.  Referring to Figure \ref{diagram}, we see that $q_{\mathrm{s}}=( t_{\mathrm{s}},\chi) $ and $q_{\ell}=\left( t_{\ell},\chi \right) $.

In \eqref{eqt1} $t_{\mathrm{s}}$ is implicitly defined as a function of $(\tau ,\chi)$, and similarly in \eqref{eqt1b} $t_{\ell}$ is implicitly defined as a function of $( \tau ,\chi) $. In this section and the next, it will be convenient to regard not only  $t_{\mathrm{s}}$ and $t_{\ell}$ as functions of $(\tau ,\chi)$, but also the four relative velocities.  However, it is important to recognize that in this context $\chi$ is a parameter that labels a comoving test particle (with fixed coordinate $\chi$), and $\tau$ is the time of observation by the central observer $\beta $.  The relative velocities are vectors in the tangent space of the point $p=(\tau,0)$ for test particles with coordinates $(t_{\mathrm{s}}, \chi)$ in the case of the Fermi and kinematic relative velocities, and with coordinates $(t_{\ell}, \chi)$ in the case of the astrometric and spectroscopic relative velocities. Since all the velocities are proportional to $\mathcal{S}_p$ and in the same direction, we will find expressions only for the moduli of the relative velocities.

\begin{remark}
\label{remretardedcomoving}
For the purpose of making retarded comparisons of comoving test particles, observe that $t_{\mathrm{s}}\left(\tau ^*,\chi \right) =t_{\ell }\left(\tau ,\chi \right)$ follows from Remark \ref{remretarded}.
\end{remark}

\subsection{Kinematic and Fermi relative velocities of comoving test particles}

Applying \eqref{vkingeneral} with constant $\chi$, we obtain
\begin{equation}
\label{eqvkin}
\| v_{\mathrm{kin}}\| =\sqrt{1-\frac{a^2(t_{\mathrm{s}})}{a^2(\tau )}}.
\end{equation}

Adapting \eqref{eqrho} to the notation of this section, the Fermi distance from $p=(\tau ,0)$ to $q_{\mathrm{s}}=(t_{\mathrm{s}}(\tau ,\chi ),\chi )$ may be expressed as a function of $(\tau ,\chi )$ as
\begin{equation}
\label{rhotauchi}
\rho=\rho (\tau ,\chi )=\int _{t_{\mathrm{s}}(\tau ,\chi )}^{\tau } \frac{a(\bar{\tau })}{\sqrt{a^2(\tau )-a^2(\bar{\tau })}}\,\textrm{d}\bar{\tau }.
\end{equation}
Then, from \eqref{fermidefinition2},
\begin{equation}
\label{vfermirho}
\| v_{\mathrm{Fermi}}\| =\frac{\partial}{\partial \tau } \int _{t_{\mathrm{s}}(\tau ,\chi )}^{\tau } \frac{a(\bar{\tau })}{\sqrt{a^2(\tau )-a^2(\bar{\tau })}}\,\textrm{d}\bar{\tau }.
\end{equation}


\subsection{Spectroscopic and astrometric relative velocities of comoving test particles}

Applying \eqref{vspecgeneral2} with constant $\chi$ gives
\begin{equation}
\label{eqvspec}
\| v_{\mathrm{spec}}\| =\frac{a^2(\tau )-a^2(t_{\ell})}{a^2(\tau )+a^2(t_{\ell})}.
\end{equation}
Similarly, with $\dot{\chi}=0$, \eqref{vastgeneral3} becomes
\begin{equation}
\label{eqvast}
\| v_{\mathrm{ast}}\| =1-\delta \frac{\dot{a}(\tau )}{a(\tau )}-\frac{a^2(t_{\ell})}{a^2(\tau )},
\end{equation}
where $\delta $ is the affine distance from $p=(\tau ,0)$ to $q_{\ell}=(t_{\ell}(\tau ,\chi ),\chi )$, and (see \eqref{delta}) it can be expressed as a function of $(\tau ,\chi )$,
\begin{equation}
\label{deltatauchi}
\delta=\delta (\tau ,\chi )=\int _{t_{\ell}(\tau ,\chi )}^{\tau} \frac{a(\bar{\tau })}{a(\tau ) }\,\textrm{d}\bar{\tau }.
\end{equation}

The astrometric speed has alternative formulations.  The cosmological redshift of a comoving test particle with respect to a central observer is well known and also follows from \eqref{taudot}:
\begin{equation}
\label{zshift}
1+z=\frac{\nu '}{\nu }=\frac{a(\tau )}{a(t_{\ell})}.
\end{equation}
Combining \eqref{eqvspec} and \eqref{zshift} in \eqref{eqvast}, we have the following result analogous to Hubble's Law for the Hubble speed.

\begin{proposition}\label{hubble}
For a test particle comoving with the Hubble flow and with redshift $z$, the astrometric relative speed relative to the central observer is given by
\begin{equation*}
\| v_{\mathrm{ast}}\| =1-H \delta -\frac{1-\| v_{\mathrm{spec}}\| }{1+\| v_{\mathrm{spec}}\| }=1- H \delta -\frac{1}{\left( 1+z\right) ^2}\,,
\end{equation*}
where $\| v_{\mathrm{spec}}\|$ is the spectroscopic relative speed and $H=\dot{a}(\tau)/a(\tau)$ is the Hubble parameter.
\end{proposition}

\section{Comoving test particles in particular spacetimes}
\label{examples}

In this section, we find explicit expressions for the relative velocities of comoving test particles for particular Robertson-Walker spacetimes. We consider the Milne universe, the de Sitter universe, and Robertson-Walker cosmologies with scale factors following a power law, i.e., $a(t)=t^{\alpha}$, where $0<\alpha<1$, including, in particular, the radiation-dominated universe and the matter-dominated universe. As in the previous section, we continue to express $t_{\mathrm{s}}$ and  $t_{\ell}$ and the moduli of the four relative velocities as functions of $\left( \tau ,\chi \right) $. In fact for all of the examples discussed in this section, this dependence is exclusively through a parameter $v$ given by
\begin{equation}
\label{hubblespeed}
v=v(\tau ,\chi ):=\dot{a}(\tau )\chi ,
\end{equation}
where the overdot represents differentiation with respect to $\tau $. However, this is not true in general, e.g. if $a(t)=t^2+t$.

\begin{remark}
\label{remhubblespeed}
The expression \eqref{hubblespeed} for $v$ is the Hubble speed of a comoving test particle with curvature-normalized coordinates $(\tau ,\chi )$.  However, it is important to recognize that the relative velocities that we calculate in this section, as functions of $v$, are those of test particles located at different spacetime points, $(t_{\mathrm{s}},\chi)$ and  $(t_{\ell}, \chi)$.
\end{remark}

Recall that for ``retarded comparisons,'' defined in Section \ref{notation}, it is necessary to calculate the kinematic and Fermi relative velocities at $p^*=(\tau ^*,0)$, which will be expressed in terms of the parameters $(\tau ^*,\chi)$ instead of $(\tau, \chi)$. For this purpose, we define $v^*:=v(\tau ^*,\chi )$.

\subsection{The Milne universe}

The Milne universe may be identified as the forward light cone in Minkowski spacetime, foliated by negatively curved hyperboloids orthogonal to the time axis. As the four relative velocities in Minkowski spacetime were previously found in \cite{Bolos07}, we include this example only for purposes of illustration of techniques presented in this paper.

For this spacetime, $a(t)=t$ and $k=-1$, but our results are valid for any $k$. From \eqref{hubblespeed}
\begin{equation}
\label{eqvHmu}
v=v(\tau ,\chi )=\chi ,
\end{equation}
which does not depend on $\tau $.

\subsubsection{Spacelike simultaneity in the Milne universe}

By \eqref{eqchiSmax}, $\chi _{\mathrm{smax}}(\tau )=\int _{0}^{\tau } \frac{\tau }{\bar{\tau }}\frac{1}{\sqrt{\tau ^2-\bar{\tau }^2}}\,\textrm{d}\bar{\tau }=+\infty $,
and hence, by \eqref{eqvHmu}, $v$ has no upper bound in the framework of spacelike simultaneity.  From \eqref{eqt1}
\begin{equation}
\label{eq:tsmilne}
\int _{t_{\mathrm{s}}}^{\tau } \frac{\tau }{\bar{\tau }}\frac{1}{\sqrt{\tau ^2-\bar{\tau }^2}}\,\textrm{d}\bar{\tau }=\chi
\quad \Longrightarrow \quad
t_{\mathrm{s}}(\tau ,\chi )=\frac{\tau }{\cosh \chi },
\end{equation}
and so by \eqref{rhotauchi}
\begin{equation}
\label{eq:rhomilne}
\rho (\tau ,\chi )=\int _{t_{\mathrm{s}}(\tau ,\chi )}^{\tau } \frac{\bar{\tau }}{\sqrt{\tau ^2-\bar{\tau }^2}}\,\textrm{d}\bar{\tau }
=\tau \tanh \chi < \tau.
\end{equation}

From \eqref{vfermirho},
\begin{equation}
\label{eq:vfermimilne}
\| v_{\mathrm{Fermi}}\| =\frac{\partial \rho }{\partial \tau }=\tanh \chi =\tanh v =\| v_{\mathrm{kin}}\|=\frac{\rho}{\tau}.
\end{equation}

Applying \eqref{eqvkin} gives
\begin{equation}
\label{eq:vkinmilne}
\| v_{\mathrm{kin}}\| =\sqrt{1-\left( \frac{t_{\mathrm{s}}}{\tau }\right) ^2}
=\tanh \chi =\tanh v=\frac{\rho}{\tau}.
\end{equation}

Thus, $\| v_{\mathrm{kin}}\|=\| v_{\mathrm{Fermi}}\|<1$.

\subsubsection{Lightlike simultaneity in the Milne universe}

By \eqref{eqchiLmax}, $\chi _{\ell \mathrm{max}}(\tau )=\int _{0}^{\tau } \frac{1}{\bar{\tau }}\,\textrm{d}\bar{\tau }=+\infty $,
and hence, by \eqref{eqvHmu}, we have that $v$ has no upper bound in the framework of lightlike simultaneity. By \eqref{eqt1b}
\begin{equation}
\label{eq:tlmilne}
\int _{t_{\ell }}^{\tau } \frac{1}{\bar{\tau }}\,\textrm{d}\bar{\tau }=\chi
\quad \Longrightarrow \quad
t_{\ell }(\tau ,\chi )=\tau e^{-\chi }.
\end{equation}
So, it follows from \eqref{deltatauchi} that
\begin{equation}
\label{eq:deltamilne}
\delta (\tau ,\chi )=\int _{t_{\ell }(\tau ,\chi )}^{\tau } \frac{\bar{\tau }}{\tau }\,\textrm{d}\bar{\tau }=\tau e^{-\chi }\sinh \chi.
\end{equation}
Then, from \eqref{eqvast}
\begin{equation}
\label{eq:vastmilne}
\| v_{\mathrm{ast}}\| =\frac{\partial \delta }{\partial \tau }= e^{-\chi }\sinh \chi =e^{-v}\sinh v = \frac{\delta}{\tau}.
\end{equation}

Similarly, \eqref{eqvspec} gives
\begin{equation}
\label{eq:vspecmilne}
\| v_{\mathrm{spec}}\| =\frac{\tau ^2-t_{\ell }^2}{\tau ^2+t_{\ell }^2}=\tanh \chi =\tanh v =\frac{\delta /\tau }{1-\delta /\tau },
\end{equation}
and it follows from \eqref{eq:deltamilne} that $\delta /\tau \in[ 0,1/2)$.

From Remark \ref{remretardedcomoving}, \eqref{eq:tsmilne} and \eqref{eq:tlmilne}, we have $\tau ^*=\tau e^{-\chi }\cosh \chi $, but for the Milne universe, a retarded comparison is equivalent to an instant comparison, because   $v^*=v$, and the relative velocities do not depend on $\tau$.

\subsection{The de Sitter universe}

For this spacetime, $a(t)=e^{H_0 t}$, where $H_0>0$, and $k=0$, but our results are valid for any $k$. We suppose that $t>0$, but we could also consider non-positive values for $t$. From \eqref{hubblespeed}
\begin{equation}
\label{eqvHdsu}
v=v(\tau ,\chi ) =H_0 e^{H_0 \tau }\chi.
\end{equation}

\subsubsection{Spacelike simultaneity for the de Sitter universe}

By \eqref{eqchiSmax}, $\chi<\chi _{\mathrm{smax}}(\tau )=\int _{0}^{\tau } \frac{e^{H_0 (\tau -\bar{\tau })}}{\sqrt{e^{2H_0 \tau }-e^{2H_0 \bar{\tau }}}}\,\textrm{d}\bar{\tau }=\frac{1}{H_0}\sqrt{1-e^{-2H_0 \tau }}$
for any point $(t, \chi)$ on the Fermi space slice at proper time $\tau$ of the central observer. Thus, by \eqref{eqvHdsu}, $0<v(\tau ,\cdot )<\sqrt{e^{2H_0 \tau }-1}$ and $v(\cdot ,\chi )>\frac{H_0\chi }{\sqrt{1-\left( H_0 \chi \right) ^2}}$.

By \eqref{eqt1}
\begin{equation}
\label{eq:tsdsu}
\int _{t_{\mathrm{s}}}^{\tau } \frac{e^{H_0 (\tau -\bar{\tau })}}{\sqrt{e^{2H_0 \tau }-e^{2H_0 \bar{\tau }}}}\,\textrm{d}\bar{\tau }=\chi
\quad \Longrightarrow \quad
t_{\mathrm{s}}(\tau ,\chi )=\tau -\frac{1}{2H_0}\ln \left( 1+v^2\right) ,
\end{equation}
and thus from \eqref{rhotauchi}
\begin{equation}
\label{eq:rhodsu}
\rho (\tau ,\chi )=\int _{t_{\mathrm{s}}(\tau ,\chi )}^{\tau } \frac{e^{H_0 \bar{\tau }}}{\sqrt{e^{2H_0 \tau }-e^{2H_0 \bar{\tau }}}}\,\textrm{d}\bar{\tau }
=\frac{1}{H_0}\arccos \left( \frac{1}{\sqrt{1+v^2}}\right).
\end{equation}
Then, from \eqref{vfermirho}
\begin{equation}
\label{eq:vfermidsu}
\| v_{\mathrm{Fermi}}\| =\frac{\partial \rho }{\partial \tau }=\frac{v}{1+v^2}=\sin \left( H_0\rho \right) \cos \left( H_0\rho \right).
\end{equation}
Applying \eqref{eqvkin} and \eqref{eq:tsdsu} gives
\begin{equation}
\label{eq:vkindsu}
\| v_{\mathrm{kin}}\| =\sqrt{1-\left( \frac{e^{H_0 t_{\mathrm{s}}}}{e^{H_0 \tau }}\right) ^2}=\frac{v}{\sqrt{1+v^2}}=\sin \left( H_0\rho \right),
\end{equation}
where $H_0 \rho \in \left[ 0,\pi /2\right)$ follows from \eqref{eq:rhodsu}.

\subsubsection{Lightlike simultaneity for the de Sitter universe}

By \eqref{eqchiLmax}, $\chi<\chi _{\ell \mathrm{max}}(\tau )=\int _{0}^{\tau } \frac{1}{e^{H_0 \bar{\tau }}}\,\textrm{d}\bar{\tau }=\frac{1}{H_0}\left( 1-e^{-H_0 \tau }\right) $
for any point $(t, \chi)$ on the past-pointing horismos $E^-_p$ (at proper time $\tau$ of the central observer). Thus, from \eqref{eqvHdsu}, $0<v(\tau ,\cdot )<e^{H_0 \tau }-1$ and $v(\cdot ,\chi )>\frac{H_0\chi }{1-H_0 \chi }$.

\begin{figure}[tbp]
\begin{center}
\includegraphics[width=0.95\textwidth]{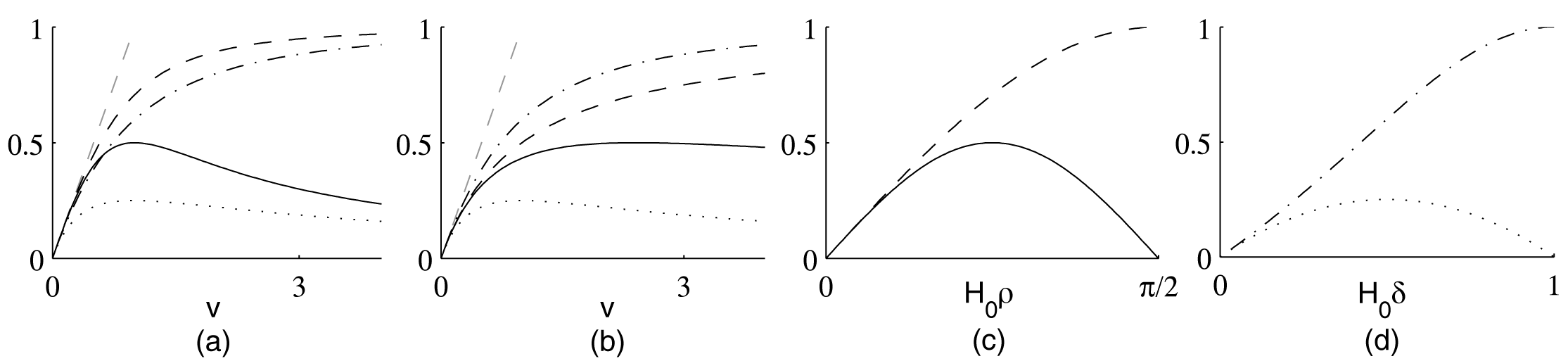}
\end{center}
\caption{Moduli of kinematic (dashed), Fermi (solid), spectroscopic (dot-dashed) and astrometric (dotted) relative velocities in the de Sitter universe with scale factor $a(t)=e^{H_0 t}$. (a): instant comparison with respect to the function $v=H_0e^{H_0\tau }\chi $; since $v$ can be interpreted as a ``Hubble speed'' (see Remark \ref{remhubblespeed}), it is represented in dashed grey in order to compare. (b): retarded comparison. (c): kinematic and Fermi velocities with respect to $H_0\rho $. (d): spectroscopic and astrometric velocities with respect to $H_0\delta $.}
\label{vdesitter_all}
\end{figure}

By \eqref{eqt1b}
\begin{equation}
\label{eq:tldsu}
\int _{t_{\ell }}^{\tau } \frac{1}{e^{H_0 \bar{\tau }}}\,\textrm{d}\bar{\tau }=\chi
\quad \Longrightarrow \quad
t_{\ell }(\tau ,\chi )=\tau -\frac{1}{H_0}\ln \left( 1+v\right).
\end{equation}
So, from \eqref{deltatauchi}
\begin{equation}
\label{eq:deltadsu}
\delta (\tau ,\chi )=\int _{t_{\ell }(\tau ,\chi )}^{\tau } \frac{e^{H_0 \bar{\tau }}}{e^{H_0 \tau }}\,\textrm{d}\bar{\tau }
=\frac{1}{H_0}\frac{v}{1+v}
\quad \Longrightarrow \quad
v=\frac{H_0 \delta}{1-H_0 \delta}.
\end{equation}
Then, by \eqref{eqvast}
\begin{equation}
\label{eq:vastdsu}
\| v_{\mathrm{ast}}\| =\frac{\partial \delta }{\partial \tau }= \frac{v}{\left( 1+v\right) ^2}=H_0 \delta \left( 1-H_0 \delta \right).
\end{equation}
Applying \eqref{eqvspec} and \eqref{eq:tldsu} gives,
\begin{equation}
\label{eq:vspecdsu}
\| v_{\mathrm{spec}}\| =\frac{e^{2H_0 \tau }-e^{2H_0 t_{\ell }}}{e^{2H_0 \tau }+e^{2H_0 t_{\ell }}}=\frac{\left( 1+v\right) ^2-1}{\left( 1+v\right) ^2+1}=\frac{1-\left( 1-H_0 \delta \right) ^2}{1+\left( 1-H_0 \delta \right) ^2},
\end{equation}
where $H_0 \delta \in \left[ 0,1\right)$ from \eqref{eq:deltadsu}.

Using Remark \ref{remretardedcomoving}, we can make retarded comparisons. From \eqref{eq:tsdsu} and \eqref{eq:tldsu},
\begin{equation}
\label{eq:vretdsu}
\tau ^*=\tau -\frac{1}{2H_0}\ln \left( 1+2v\right)
\quad \Longrightarrow \quad
v^*=\frac{v}{\sqrt{1+2v}}.
\end{equation}
So,
\begin{equation}
\label{eq:vkinfermiretdsu}
\| v_{\mathrm{kin}}^*\| =\frac{v^*}{\sqrt{1+v^{*2}}}=\frac{v}{1+v}\quad ;\quad
\| v_{\mathrm{Fermi}}^*\| =\frac{v^*}{1+v^{*2}}=\frac{v\sqrt{1+2v}}{\left( 1+v\right) ^2}.
\end{equation}
Observe that all of the relative velocities for the de Sitter universe are independent of $\tau $ (see Figure \ref{vdesitter_all}).

\subsection{Power scale factor}

In this section, we consider a power scale factor of the form $a(t)=t^\alpha $ with $0<\alpha <1$. There are some important particular cases, as $\alpha =1/3$, $\alpha =1/2$ (radiation-dominated universe), or $\alpha =2/3$ (matter-dominated universe). From \eqref{hubblespeed},
\begin{equation}
\label{eqvHpsf}
v=v(\tau ,\chi )=\frac{\alpha }{\tau ^{1-\alpha }}\chi .
\end{equation}

\subsubsection{Spacelike simultaneity for $a(t)=t^\alpha$}

It follows from \eqref{eqchiSmax} that $\chi<\chi _{\mathrm{smax}}(\tau )=\int _{0}^{\tau } \frac{\tau ^{\alpha }}{\bar{\tau }^{\alpha }}\frac{1}{\sqrt{\tau ^{2\alpha }-\bar{\tau }^{2\alpha }}}\,\textrm{d}\bar{\tau }=\frac{\tau ^{1-\alpha}}{1-\alpha}C_{\alpha}$
for any point $(t, \chi)$ on the Fermi space slice at proper time $\tau$ of the central observer, where $C_{\alpha }:=\frac{\sqrt{\pi }\,\Gamma \left( \frac{1+\alpha }{2\alpha }\right) }{\Gamma \left( \frac{1}{2\alpha }\right) }$. Hence, by \eqref{eqvHpsf}, we have that $0<v<v_{\mathrm{smax}}$ where,
\begin{equation}
\label{eq:vsmaxpsf}
v_{\mathrm{smax}}:=v\left( \tau , \chi _{\mathrm{smax}}(\tau )\right) =\frac{\alpha}{1-\alpha}C_{\alpha }.
\end{equation}
We have that $v_{\mathrm{smax}}$ is unbounded and is an increasing function of $\alpha $.  It reaches $1$ at $\alpha =1/3$.

By \eqref{eqt1}, $\int _{t_{\mathrm{s}}}^{\tau } \frac{\tau ^{\alpha }}{\bar{\tau }^{\alpha }}\frac{1}{\sqrt{\tau ^{2\alpha }-\bar{\tau }^{2\alpha }}}\,\textrm{d}\bar{\tau }=\chi $, and then
\begin{equation}
\label{eq:tspsf}
\left( \frac{t_{\mathrm{s}}}{\tau }\right) ^{1-\alpha } \,\!_2F_1\left( \frac{1}{2},\frac{1-\alpha }{2\alpha };\frac{1+\alpha }{2\alpha };\left( \frac{t_{\mathrm{s}}}{\tau }\right) ^{2\alpha }\right) = C_{\alpha }-\frac{1-\alpha }{\alpha }v,
\end{equation}
where $ _2F_1(\cdot ,\cdot ;\cdot ;\cdot )$ is the Gauss hypergeometric function. Define the function $F_{\alpha }(z):=z^{\frac{1-\alpha }{\alpha }} \, _2F_1\left( \frac{1}{2},\frac{1-\alpha }{2\alpha };\frac{1+\alpha }{2\alpha };z^2\right) $
where $0<z<1$ and $0<\alpha <1$. It is bijective and by \eqref{eq:tspsf}
\begin{equation}
\label{eqt1Galpha}
t_{\mathrm{s}}(\tau ,\chi ) =G_{\alpha }(v) \tau,
\end{equation}
with $G_{\alpha }(v):=\left( F_{\alpha }^{-1}\left( C_{\alpha }-\frac{1-\alpha }{\alpha }v\right) \right)^{1/\alpha }$,
where the superscript $^{-1}$ denotes the inverse function. From \eqref{rhotauchi} and \eqref{eqt1Galpha}
\begin{equation}
\label{eq:rhopsf}
\rho (\tau ,\chi ) = \int _{t_{\mathrm{s}}(\tau ,\chi )}^{\tau } \frac{\bar{\tau }^{\alpha }}{\sqrt{\tau ^{2\alpha }-\bar{\tau }^{2\alpha }}}\,\textrm{d}\bar{\tau }= J_{\alpha }(v)\tau ,
\end{equation}
where $J_{\alpha }(v):=C_{\alpha }-\frac{G_{\alpha }^{1+\alpha }(v)}{1+\alpha } \, _2F_1\left( \frac{1}{2},\frac{1+\alpha }{2\alpha };\frac{1+3\alpha }{2\alpha };G_{\alpha }^{2\alpha }(v)\right) $.
Setting $t_{\mathrm{s}}(\tau ,\chi )=0$ in \eqref{eq:rhopsf} results in $\rho(\tau, \chi _{\mathrm{smax}}) = C_{\alpha } \tau$, the radius of the Fermi space slice at the central observer's proper time $\tau$, which was also found in \cite{Klein11}.

By \eqref{vfermirho} and \eqref{eq:rhopsf}
\begin{equation}
\label{eqvfermitalpha}
\| v_{\mathrm{Fermi}}\| =\frac{\partial \rho }{\partial \tau }=J_{\alpha }(v)-(1-\alpha )J'_{\alpha }(v)v.
\end{equation}
Differentiating $G_{\alpha }$ with respect to $v$, we obtain $G'_{\alpha }=-\frac{1}{\alpha }G_{\alpha }^{\alpha }\sqrt{1-G_{\alpha }^{2\alpha }}$,
which can also be found by differentiating \eqref{eqt1} with respect to $v$, taking into account that $\chi =\frac{v}{\dot{a}(t_{\mathrm{s}})}$ and \eqref{eqt1Galpha}. Integrating this differential equation and using the initial condition, $G_{\alpha }(0)=1$, and the expression of $J_{\alpha }$, gives
\begin{equation}
\label{eqJalpha2}
J_{\alpha }(v)=G_{\alpha }^{1-\alpha }(v)\sqrt{1-G_{\alpha }^{2\alpha }(v)}+\frac{1-\alpha }{\alpha }v\quad \Longrightarrow \quad
J'_{\alpha }(v)=\frac{1}{\alpha}G_{\alpha }^{2\alpha }(v).
\end{equation}
Combining \eqref{eqvkin} and \eqref{eqt1Galpha}
along with \eqref{eqJalpha2}, \eqref{eq:rhopsf}, and \eqref{eqvfermitalpha}, we have,
\begin{proposition}\label{powerVFermi}
The kinematic and Fermi speeds of a comoving test particle relative to a comoving central observer in a Robertson-Walker cosmology with scale factor $a(t)=t^{\alpha}$ and $0<\alpha<1$ are given by
\begin{equation}
\label{powerVkinetic}
\| v_{\mathrm{kin}}\| =\sqrt{1-\frac{t_{\mathrm{s}}^{2\alpha }}{\tau ^{2\alpha }}}= \sqrt{1-G_{\alpha }^{2\alpha }(v)},
\end{equation}

\begin{equation}
\label{eq:vFermi3psf}
\| v_{\mathrm{Fermi}}\| =G_{\alpha }^{1-\alpha }(v)\sqrt{1-G_{\alpha }^{2\alpha }(v) }+\frac{1-\alpha }{\alpha }\left( 1-G_{\alpha }^{2\alpha }(v)\right) v=\frac{\rho }{\tau }-\frac{1-\alpha}{\alpha } G_{\alpha }^{2\alpha }(v)v.
\end{equation}
\end{proposition}

\begin{remark}
\label{rem:newpsf1}
From \eqref{eq:rhopsf} and since $J_{\alpha }(v)$ is bijective for $0<v<v_{\mathrm{smax}}$, we have $v=J_{\alpha }^{-1}\left( \frac{\rho }{\tau }\right) $,
where the superscript $^{-1}$ denotes the inverse function. So, by \eqref{powerVkinetic} and \eqref{eqvfermitalpha} (or Proposition \ref{powerVFermi}) we can deduce expressions for $\| v_{\mathrm{kin}}\|$ and $\| v_{\mathrm{Fermi}}\|$ in terms of $\rho /\tau \in \left[ 0,C_{\alpha }\right) $.
\end{remark}

\begin{remark}
An alternative expression to \eqref{eq:vFermi3psf} was given in \cite{Klein11},
\begin{equation}
\label{distanceovertime}
\| v_{\mathrm{Fermi}}\| = \frac{\rho }{\tau }-\frac{1-\alpha }{\alpha }\frac{1}{2\sigma _0}\int_1^{\sigma _0}\frac{1}{\sigma ^{\frac{1}{2\alpha }}\sqrt{\sigma -1}}\,\textrm{d}\sigma ,
\end{equation}
where $\sigma_{0}=(a(\tau)/a(t_{\mathrm{s}}))^2=\left( \tau/t_{\mathrm{s}}\right) ^{2\alpha }=1/G_{\alpha }^{2\alpha }(v)$.
Using \eqref{eq:tspsf}, it can be shown that $v=\frac{1}{2}\int_1^{\sigma _0}\frac{1}{\sigma ^{\frac{1}{2\alpha }}\sqrt{\sigma -1}}\,\textrm{d}\sigma $, and so \eqref{eq:vFermi3psf} and \eqref{distanceovertime} are equivalent.
\end{remark}

\begin{figure}[tp]
\begin{center}
\includegraphics[width=0.6\textwidth]{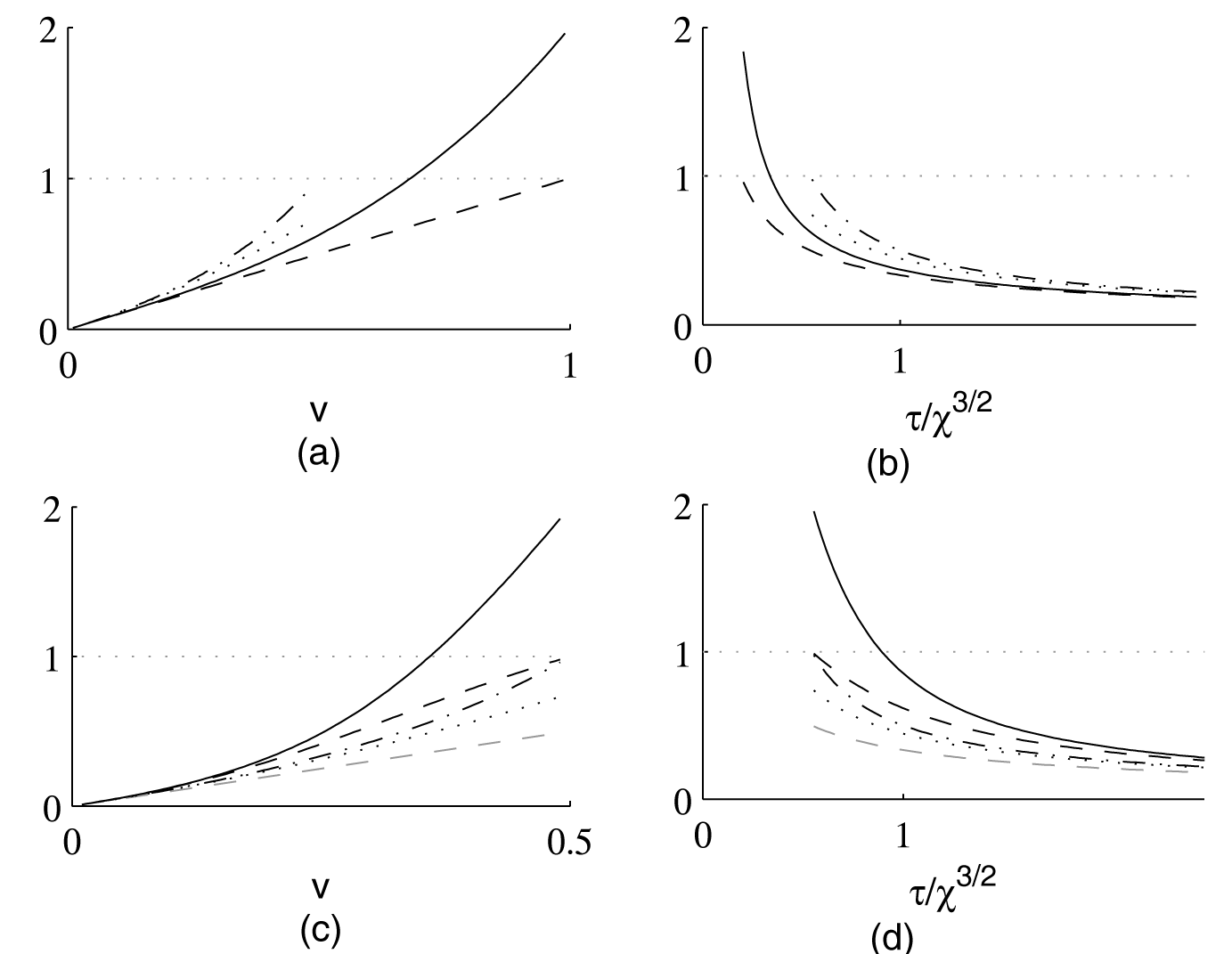}
\end{center}
\caption{Instant ((a), (b)) and retarded ((c), (d)) comparison of the moduli of kinematic (dashed), Fermi (solid), spectroscopic (dot-dashed) and astrometric (dotted) relative velocities with respect to the function $v=\chi /3\tau ^{2/3}$ ((a), (c)) and with respect to $\tau /\chi ^{3/2}$ ((b), (d)), in a universe with power scale factor $a(t)=t^{1/3}$. Since $v$ can be interpreted as a ``Hubble speed'' (see Remark \ref{remhubblespeed}), it is represented in dashed grey in order to compare, but it coincides with the kinematic relative velocity in the instant comparison.}
\label{vpsf13}
\end{figure}

We turn now to examples for particular values of $\alpha$.

\begin{example}
\label{example1}
If $\alpha =1/3$ then, from \eqref{eq:vsmaxpsf} we have $v_{\mathrm{smax}}=1$. From \eqref{eq:tspsf} we get $t_{\mathrm{s}}(\tau ,\chi ) =(1-v^2)^{3/2}\tau $, and by \eqref{eq:rhopsf} we have $\rho (\tau ,\chi ) =v(3-v^2)\tau $. So, by  \eqref{fermidefinition2} and Proposition \ref{powerVFermi} we find
\begin{equation}
\label{psf13vkinfermi}
\| v_{\mathrm{kin}}\| =v\quad ;\quad
\| v_{\mathrm{Fermi}}\| =v(1+v^2).
\end{equation}
As indicated in Remark \ref{rem:newpsf1} we find $v=2\sin\left( \frac{1}{3}\arcsin \left( \frac{\rho /\tau }{2}\right) \right) $, and then applying \eqref{psf13vkinfermi} we can get expressions for $\| v_{\mathrm{kin}}\| $ and $\| v_{\mathrm{Fermi}}\| $ in terms of $\rho /\tau \in \left[ 0,2\right) $ (see Figure \ref{fig_vpsf_dtrt}).
\end{example}

\begin{example}
\label{example2}
For $\alpha =1/2$, i.e., for the radiation-dominated universe, \eqref{eq:vsmaxpsf} becomes $v_{\mathrm{smax}}=\pi /2$. By \eqref{eq:tspsf}, $t_{\mathrm{s}}(\tau ,\chi ) =\tau \cos ^2 v$, and from \eqref{eq:rhopsf} we find $\rho (\tau ,\chi ) =(v+\cos v \sin v)\tau $. So, by  \eqref{fermidefinition2} and Proposition \ref{powerVFermi} we find
\begin{equation}
\label{psf12vkinfermi}
\| v_{\mathrm{kin}}\| =\sin v \quad ; \quad
\| v_{\mathrm{Fermi}}\| =\left( \cos v +v \sin v \right) \sin v.
\end{equation}
As indicated in Remark \ref{rem:newpsf1}, we find that \eqref{eq:rhopsf} defines implicitly $v\left( \rho /\tau \right) $, and then, from \eqref{psf12vkinfermi} we can calculate $\| v_{\mathrm{kin}}\| $ and $\| v_{\mathrm{Fermi}}\| $ in terms of $\rho /\tau \in \left[ 0,\pi /2\right) $ (see Figure \ref{fig_vpsf_dtrt}).
\end{example}

\begin{example}
\label{example3}
For $\alpha =2/3$, the matter-dominated universe, \eqref{eq:vsmaxpsf} becomes $v_{\mathrm{smax}}=2C_{2/3}=2\frac{\sqrt{\pi }\Gamma \left( 5/4\right) }{\Gamma \left( 3/4\right) }\approx 2.62206$. From \eqref{eq:tspsf} we get $t_{\mathrm{s}}(\tau ,\chi ) =\textrm{cd}^3\left( \frac{v}{2}\,|-1\right) \tau $, and by \eqref{eq:rhopsf} we get $\rho (\tau ,\chi ) =\left( \textrm{cd}\left( \frac{v}{2}\,|-1\right) \sqrt{1-\textrm{cd}^4\left( \frac{v}{2}\,|-1\right) }+\frac{v}{2} \right) \tau $, where $\textrm{cd}\left( u\,|\,m\right) $ is a Jacobi elliptic function. So, by  \eqref{fermidefinition2} and Proposition \ref{powerVFermi} we find
\begin{equation}
\label{psf23vkinfermi}
\| v_{\mathrm{kin}}\| =\sqrt{1-\textrm{cd}^4\left( \frac{v}{2}\,|-1\right) } \quad ; \quad
\| v_{\mathrm{Fermi}}\| =\textrm{cd}\left( \frac{v}{2}\,|-1\right) \| v_{\mathrm{kin}}\|+\frac{v}{2}\| v_{\mathrm{kin}}\|^2.
\end{equation}
As in the previous two examples, equation \eqref{eq:rhopsf} defines implicitly the function $v\left( \rho /\tau \right) $, and then, from \eqref{psf23vkinfermi} we can compute $\| v_{\mathrm{kin}}\| $ and $\| v_{\mathrm{Fermi}}\| $ in terms of $\rho /\tau \in \left[ 0,C_{2/3}\right)$ (see Figure \ref{fig_vpsf_dtrt}).
\end{example}

\subsubsection{Lightlike simultaneity for $a(t)=t^\alpha$}

By \eqref{eqchiLmax}, $\chi<\chi _{\ell \mathrm{max}}(\tau )=\int _{0}^{\tau } \frac{1}{\bar{\tau }^{\alpha }}\,\textrm{d}\bar{\tau }=\frac{\tau ^{1-\alpha }}{1-\alpha }$
for any point $(t, \chi)$ on the past-pointing horismos $E^-_p$ (at proper time $\tau$ of the central observer). Thus from \eqref{eqvHpsf}, we have that $0<v<v_{\ell \mathrm{max}}$ where
\begin{equation}
\label{eq:vlmaxpsf}
v_{\ell \mathrm{max}}:=v\left( \tau , \chi _{\ell \mathrm{max}}(\tau )\right)  =\frac{\alpha }{1-\alpha }.
\end{equation}
We have that $v_{\ell \mathrm{max}}$ is increasing with $\alpha $ and it reaches $1$ at $\alpha =1/2$.

\begin{figure}[tp]
\begin{center}
\includegraphics[width=0.6\textwidth]{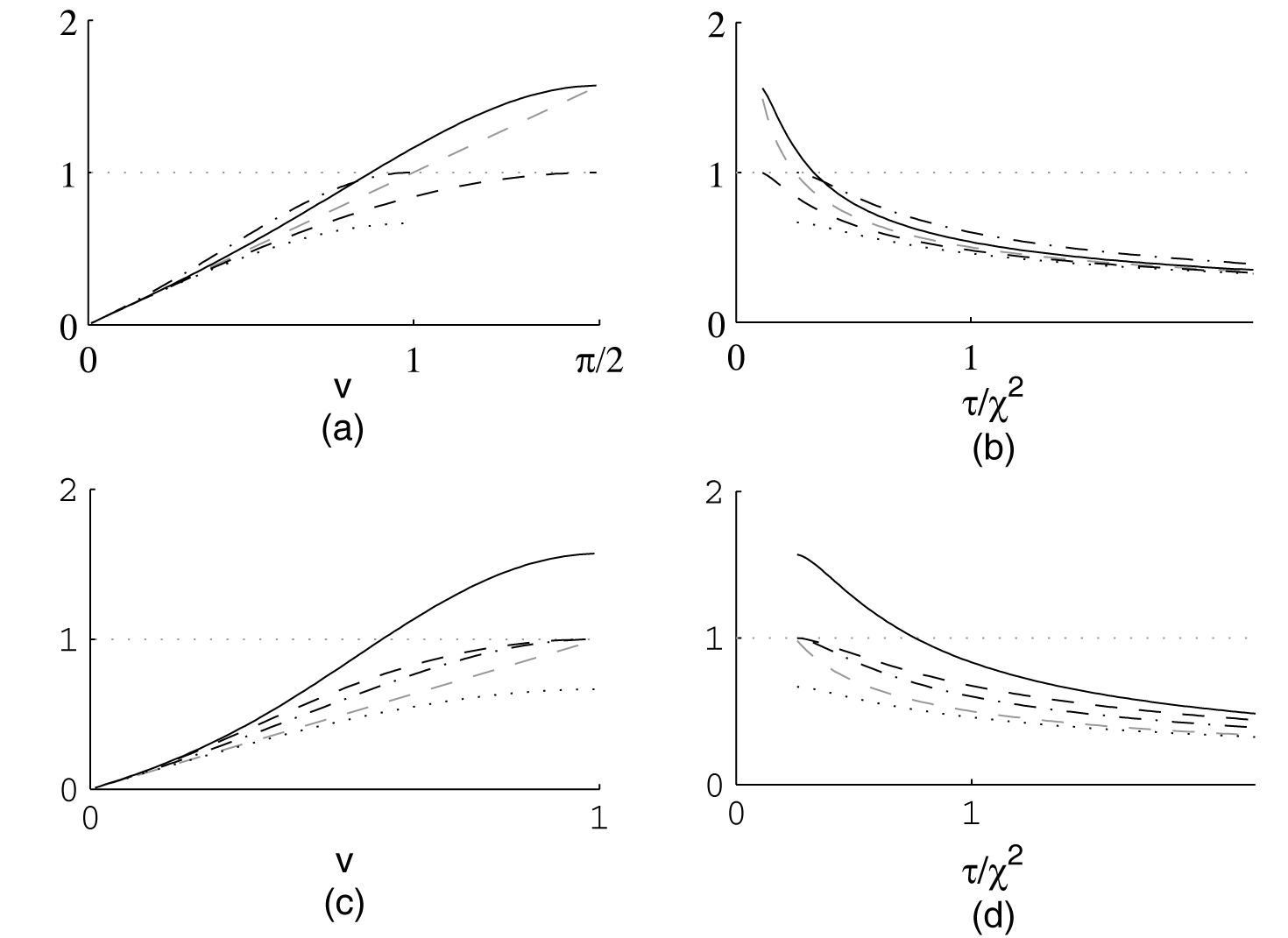}
\end{center}
\caption{Instant ((a), (b)) and retarded ((c), (d)) comparison of the moduli of kinematic (dashed), Fermi (solid), spectroscopic (dot-dashed) and astrometric (dotted) relative velocities with respect to the function $v=\chi /2\tau ^{1/2}$ ((a), (c)) and with respect to $\tau /\chi ^2$ ((b), (d)), in a universe with power scale factor $a(t)=t^{1/2}$ (radiation-dominated universe). Since $v$ can be interpreted as a ``Hubble speed'' (see Remark \ref{remhubblespeed}), it is represented in dashed grey in order to compare.}
\label{vrdu}
\end{figure}

By \eqref{eqt1b}
\begin{equation}
\label{eq:tlpsf}
\int _{t_{\ell }}^{\tau } \frac{1}{\bar{\tau }^{\alpha }}\,\textrm{d}\bar{\tau }=\chi
\quad \Longrightarrow \quad
t_{\ell }(\tau ,\chi )=\left( \tau ^{1-\alpha }-(1-\alpha )\chi \right) ^{\frac{1}{1-\alpha }}=\left( 1-\frac{1-\alpha }{\alpha }v\right) ^{\frac{1}{1-\alpha }}\tau .
\end{equation}
So, from \eqref{deltatauchi}
\begin{equation}
\label{eq:deltapsf}
\delta (\tau ,\chi )=\int _{t_{\ell }(\tau ,\chi )}^{\tau } \frac{\bar{\tau }^{\alpha }}{\tau ^{\alpha }}\,\textrm{d}\bar{\tau }
=\frac{1}{1+\alpha }\left( 1-\left( 1-\frac{1-\alpha }{\alpha }v \right) ^{\frac{1+\alpha }{1-\alpha }}\right) \tau .
\end{equation}
Now using \eqref{vastoptical}, \eqref{eqvspec} and \eqref{eqvast} together with \eqref{eq:tlpsf} and  \eqref{eq:deltapsf}, we have,
\begin{proposition}\label{powerlight}
The spectroscopic and astrometric speeds of a comoving test particle relative to a comoving central observer in a Robertson-Walker cosmology with scale factor $a(t)=t^{\alpha}$ and $0<\alpha<1$ are given by
\begin{equation}
\label{eq:vspecastpsf}
\| v_{\mathrm{spec}}\| =\frac{1-\left( 1-\frac{1-\alpha }{\alpha }v\right) ^{\frac{2\alpha }{1-\alpha }}}{1+\left( 1-\frac{1-\alpha }{\alpha }v\right) ^{\frac{2\alpha }{1-\alpha }}}
\quad ;\quad \| v_{\mathrm{ast}}\|  =
\frac{\delta }{\tau }-\frac{1-\alpha }{\alpha }v\left( 1-\frac{1-\alpha }{\alpha }v\right) ^{\frac{2\alpha }{1-\alpha }}.
\end{equation}
\end{proposition}

\begin{remark}
\label{rem:newpsf2}
From \eqref{eq:deltapsf} we obtain
\begin{equation}
\label{eq:vdeltataupsf}
\frac{1-\alpha }{\alpha }v=1-\left( 1-\left( 1+\alpha \right) \frac{\delta }{\tau } \right) ^{\frac{1-\alpha }{1+\alpha }}.
\end{equation}
Then, from Proposition \ref{powerlight} we can express $\| v_{\mathrm{spec}}\|$ and $\| v_{\mathrm{ast}}\|$ in terms of $\delta /\tau \in \left[ 0,\frac{1}{1+\alpha }\right)$.
\end{remark}

Using Remark \ref{remretardedcomoving}, we can make retarded comparisons. From  \eqref{eqt1Galpha} and \eqref{eq:tlpsf} one can solve for $\tau ^*$. Then, with the $^*$-version of \eqref{eqvHpsf} we have that $v^*=\frac{\alpha }{\left( \tau ^*\right) ^{1-\alpha }}\chi $, and
\begin{equation}
\label{eq:vkinfermiretpsf}
\| v_{\mathrm{kin}}^*\| = \sqrt{1-G_{\alpha }^{2\alpha }(v^*)} \quad ;\quad
\| v_{\mathrm{Fermi}}^*\| = G_{\alpha }^{1-\alpha }(v^*)\| v_{\mathrm{kin}}^*\|+\frac{1-\alpha }{\alpha }\| v_{\mathrm{kin}}^*\|^2v^*.
\end{equation}
Again from \eqref{eqvHpsf}, we have $\tau =\left( \frac{\alpha }{v}\chi \right) ^{\frac{1}{1-\alpha}}$ and $\tau ^*=\left( \frac{\alpha }{v^*}\chi \right) ^{\frac{1}{1-\alpha}}$. So, from \eqref{eqt1Galpha} and \eqref{eq:tlpsf} we can solve $v^*$ directly in terms of $v$ using the equation
\begin{equation}
\label{eq:vretvpsf}
\left( \,\!_2F_1\left( \frac{1}{2},\frac{1-\alpha }{2\alpha };\frac{1+\alpha }{2\alpha };w^{\frac{2\alpha }{1-\alpha }}\right) +\frac{v}{\frac{\alpha }{1-\alpha }-v} \right) w=C_{\alpha },
\end{equation}
where $w:=\left( \frac{1}{v}-\frac{1-\alpha }{\alpha }\right) v^*$.

\begin{figure}[tp]
\begin{center}
\includegraphics[width=0.6\textwidth]{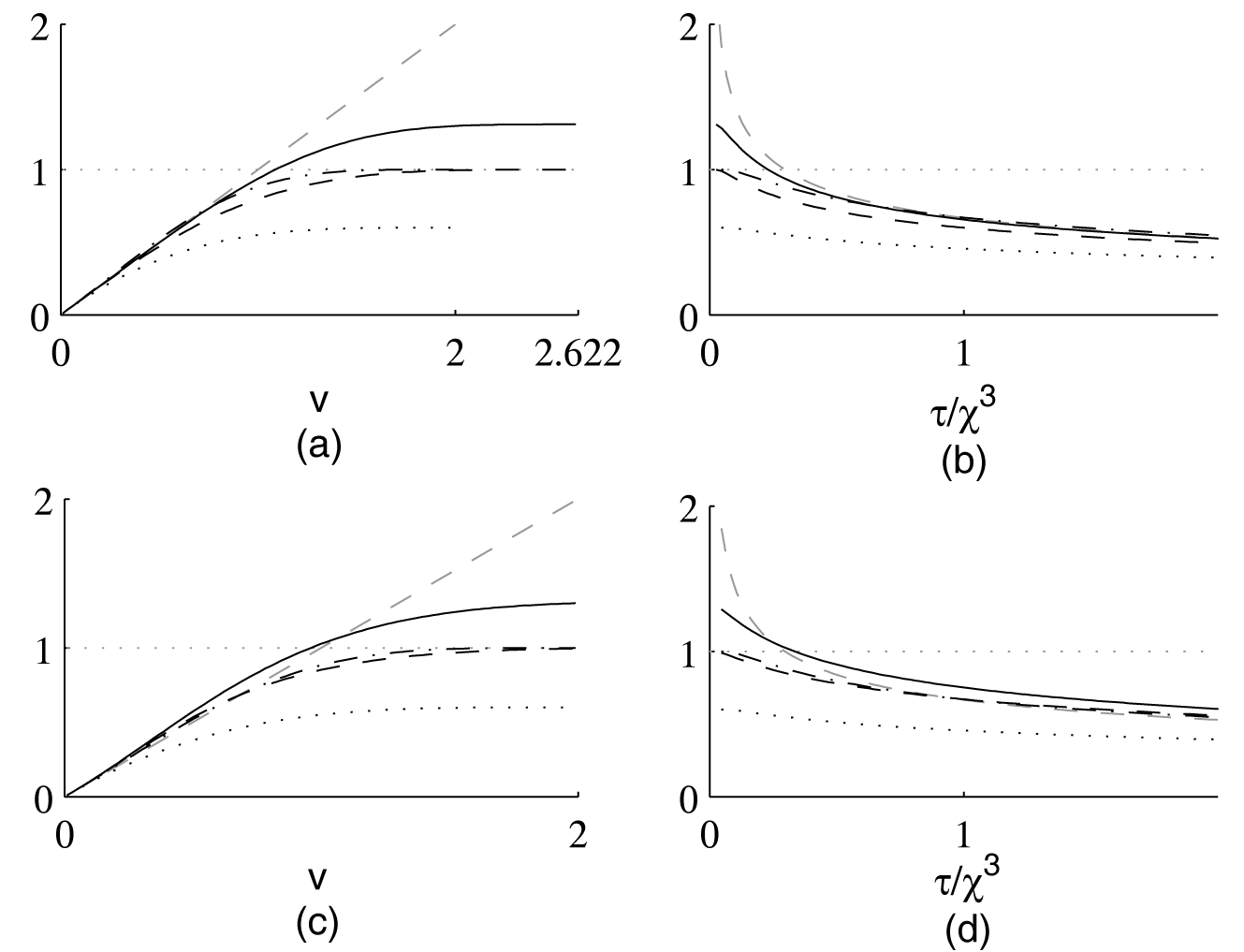}
\end{center}
\caption{Instant ((a), (b)) and retarded ((c), (d)) comparison of the moduli of kinematic (dashed), Fermi (solid), spectroscopic (dot-dashed) and astrometric (dotted) relative velocities with respect to the function $v=3\chi /2\tau^{1/3}$ ((a), (c)) and with respect to $\tau /\chi ^3$ ((b), (d)), in a universe with power scale factor $a(t)=t^{2/3}$ (matter-dominated universe). Since $v$ can be interpreted as a ``Hubble speed'' (see Remark \ref{remhubblespeed}), it is represented in dashed grey in order to compare.}
\label{vmdu}
\end{figure}

\begin{figure}[tp]
\begin{center}
\includegraphics[width=0.7\textwidth]{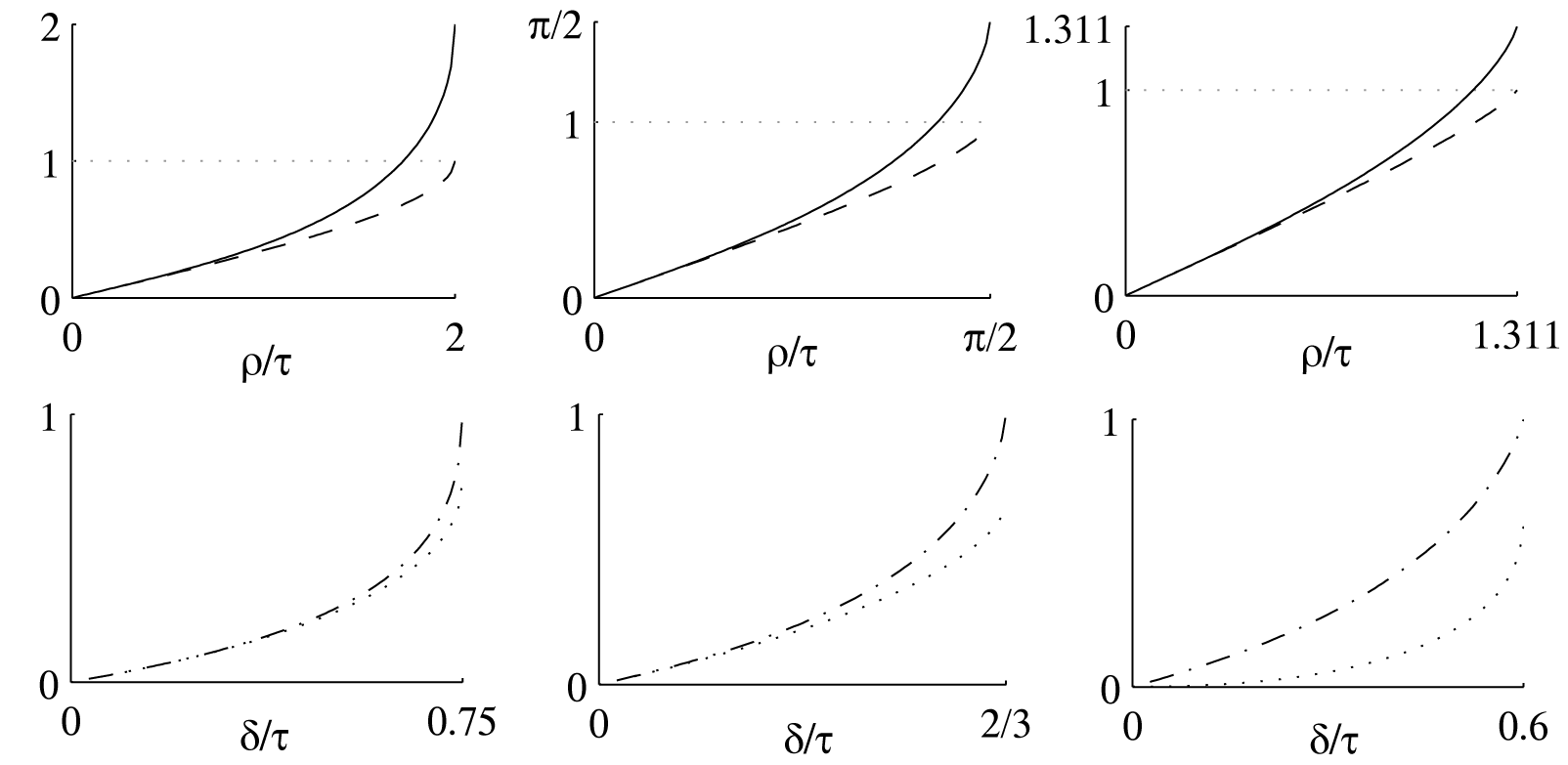}
\end{center}
\caption{Moduli of the relative velocities in a universe with power scale factor $a(t)=t^{\alpha }$ with $\alpha = 1/3$ (left), $\alpha =1/2$ (middle) and $\alpha =2/3$ (right). Top: kinematic (dashed) and Fermi (solid) velocities with respect to $\rho /\tau $. Bottom: spectroscopic (dot-dashed) and astrometric (dotted) velocities with respect to $\delta /\tau $.}
\label{fig_vpsf_dtrt}
\end{figure}

\begin{example}
If $\alpha =1/3$, then from \eqref{eq:vlmaxpsf} $v_{\ell \mathrm{max}}=1/2$. From \eqref{eq:tlpsf} we get $t_{\ell }(\tau ,\chi ) =(1-2v)^{3/2} \tau $, and by \eqref{eq:deltapsf}, $\delta (\tau ,\chi ) =3v\left( 1-v\right) \tau $. So, from Proposition \ref{powerlight},
\begin{equation}
\label{eq:vspecastpsf13}
\| v_{\mathrm{spec}}\| =\frac{v}{1-v} \quad ; \quad
\| v_{\mathrm{ast}}\| =v\left( 1+v\right) .
\end{equation}
Taking into account Remark \ref{remretardedcomoving}, from \eqref{eqt1Galpha} and \eqref{eq:tlpsf} we have $v^*=\left( \left( \frac{1}{2v}-1\right) ^2+1\right) ^{1/2}-\left( \frac{1}{2v}-1\right) $. So, $\| v_{\mathrm{kin}}^*\|$ and $\| v_{\mathrm{Fermi}}^*\|$ are given by \eqref{psf13vkinfermi}, with $v^*$ replacing $v$.

As Remark \ref{rem:newpsf2} indicates, from \eqref{eq:vdeltataupsf} or \eqref{eq:deltapsf} we obtain $2v=1-\left( 1-\frac{4}{3}\frac{\delta }{\tau }\right) ^{1/2}$ and then, from \eqref{eq:vspecastpsf13} we get $\| v_{\mathrm{spec}}\| =\frac{1-\left( 1-\frac{4}{3}\frac{\delta }{\tau }\right) ^{1/2}}{1+\left( 1-\frac{4}{3}\frac{\delta }{\tau }\right) ^{1/2}} $ and $\| v_{\mathrm{ast}}\| =1-\frac{1}{3}\frac{\delta }{\tau }-\left( 1-\frac{4}{3}\frac{\delta }{\tau }\right) ^{1/2}$, where $\delta /\tau \in \left[ 0,3/4\right) $ (see Figure \ref{fig_vpsf_dtrt}).
\end{example}

\begin{example}
For the radiation-dominated universe, with $\alpha =1/2$, \eqref{eq:vlmaxpsf} gives $v_{\ell \mathrm{max}}=1$. From \eqref{eq:tlpsf}, $t_{\ell }(\tau ,\chi ) =(1-v)^2 \tau $, and \eqref{eq:deltapsf} becomes $\delta (\tau ,\chi ) =\frac{2}{3}\left( 1-(1-v)^3\right) \tau $. So, from \eqref{eq:vspecastpsf} we obtain
\begin{equation}
\label{eq:vspecastpsf12}
\| v_{\mathrm{spec}}\| =\frac{1-\left( 1-v \right) ^2}{1+\left( 1-v \right) ^2} \quad ; \quad
\| v_{\mathrm{ast}}\| =v \left( 1-\frac{1}{3}v ^2\right) .
\end{equation}
Applying Remark \ref{remretardedcomoving}, along with \eqref{eqt1Galpha} and \eqref{eq:tlpsf}, $v^*$ is the inverse of $v=\frac{v^*}{v^*+\cos v^*} $, where $0<v^*<v_{\mathrm{smax}}=\pi /2$. So, $\| v_{\mathrm{kin}}^*\|$ and $\| v_{\mathrm{Fermi}}^*\|$ are given by \eqref{psf12vkinfermi}, with $v^*$ replacing $v$.

From Remark \ref{rem:newpsf2} and \eqref{eq:vdeltataupsf}, we get $v=1-\left( 1-\frac{3}{2}\frac{\delta }{\tau }\right) ^{1/3}$ and then \eqref{eq:vspecastpsf12} results in $\| v_{\mathrm{spec}}\| =\frac{1-\left(1-\frac{3}{2}\frac{\delta }{\tau }\right) ^{2/3}}{1+\left(1-\frac{3}{2}\frac{\delta }{\tau }\right) ^{2/3}} $ and $\| v_{\mathrm{ast}}\| =1-\frac{1}{2}\frac{\delta }{\tau }-\left(1-\frac{3}{2}\frac{\delta }{\tau }\right) ^{2/3}$, where $\delta /\tau \in \left[ 0,2/3\right) $ (see Figure \ref{fig_vpsf_dtrt}).
\end{example}

\begin{example}
For the matter-dominated universe, with $\alpha =2/3$, from \eqref{eq:vlmaxpsf} we have $v_{\ell \mathrm{max}}=2$. From \eqref{eq:tlpsf}, $t_{\ell }(\tau ,\chi ) =\left( 1-\frac{v}{2}\right) ^3 \tau $, and \eqref{eq:deltapsf} becomes $\delta (\tau ,\chi ) =\frac{3}{5}\left( 1-\left( 1-\frac{v}{2}\right) ^5\right) \tau $. So, by \eqref{eq:vspecastpsf} we obtain
\begin{equation}
\label{eq:vspecastpsf23}
\| v_{\mathrm{spec}}\| =\frac{1-\left( 1-\frac{v}{2}\right) ^4}{1+\left( 1-\frac{v}{2}\right) ^4} \quad ; \quad
\| v_{\mathrm{ast}}\| =\left( 1-\frac{v}{2}+\frac{1}{2}\left( \frac{v}{2}\right) ^3 -\frac{1}{5}\left( \frac{v}{2}\right) ^4\right) v.
\end{equation}
Taking into account Remark \ref{remretardedcomoving}, from \eqref{eqt1Galpha} and \eqref{eq:tlpsf} we obtain $v^*$ as the inverse of $v=\frac{2v^*}{v^*+2\textrm{cd}\left( \frac{\,v^*}{2}\,|-1\right) }$, where $0<v^*<v_{\mathrm{smax}}=2C_{2/3}\approx 2.62206$. So, $\| v_{\mathrm{kin}}^*\|$ and $\| v_{\mathrm{Fermi}}^*\|$ are given by \eqref{psf23vkinfermi}, considering $v^*$ instead of $v$.

From Remark \ref{rem:newpsf2} and \eqref{eq:vdeltataupsf} we get $\frac{v}{2}=1-\left( 1-\frac{5}{3}\frac{\delta }{\tau }\right) ^{1/5}$ and then \eqref{eq:vspecastpsf23} gives $\| v_{\mathrm{spec}}\| =\frac{1-\left( 1-\frac{5}{3}\frac{\delta }{\tau }\right) ^{4/5}}{1+\left( 1-\frac{5}{3}\frac{\delta }{\tau }\right) ^{4/5}}$ and $\| v_{\mathrm{ast}}\| =1-\frac{2}{3}\frac{\delta }{\tau }-2\left( 1-\frac{5}{3}\frac{\delta }{\tau }\right) ^{3/5}+\left( 1-\frac{5}{3}\frac{\delta }{\tau }\right) ^{4/5}$, where $\delta /\tau \in \left[ 0,3/5\right) $ (see Figure \ref{fig_vpsf_dtrt}).
\end{example}

\section{Functional relationships for relative velocities}
\label{sec:5}

In this section, we allow the radial motion of test particles to be non geodesic. Throughout, we assume that the scale factor $a(t)$ is smooth and increasing. We employ both Fermi coordinates and optical coordinates to derive functional relationships between the Fermi, kinematic, astrometric, and spectroscopic relative velocities of test particles. First, we relate Fermi and kinematic relative velocities to each other and spectroscopic and astrometric relative velocities to each other, and consider some geometric consequences of these relationships. Then, in Subsections \ref{secrel1} and \ref{secrel2}, we use Fermi coordinates to find retarded relationships for the spectroscopic and astrometric relative velocities respectively.

We begin with a relationship between the Fermi and kinematic relative velocities that follows immediately from \eqref{vkinfermi}:
\begin{equation}\label{compare5}
v_{\mathrm{Fermi}}=\sqrt{-g_{\tau\tau}(\tau,\rho)}\, v_{\mathrm{kin}}.
\end{equation}

Analogously, in the context of lightlike simultaneity, combining \eqref{vspecinoptical} with \eqref{vastoptical} gives
\begin{equation}\label{astro/spec}
v_{\mathrm{spec}}=\frac{1+\tilde{g}_{\tau\tau}(\tau,\delta) \pm \| v_{\mathrm{ast}}\|}{1-\tilde{g}_{\tau\tau}(\tau,\delta) \mp \| v_{\mathrm{ast}}\|}\mathcal{S}_p,
\end{equation}
where in the case that $d\delta/d\tau>0$ (i.e., in the case of a receding test particle), the positive sign in the numerator and negative sign in the denominator are chosen, and the opposite choices of signs are taken when $d\delta/d\tau<0$ (i.e., in the case of an approaching test particle).  From  \eqref{compare5} and \eqref{astro/spec} we may formulate the following proposition.

\begin{proposition}\label{findmetric}
For a Robertson-Walker spacetime with scale factor $a(t)$ that is a smooth, increasing function of $t$, the kinematic and Fermi speeds of any test particle undergoing radial motion with respect to a comoving observer determine the Fermi metric tensor element $g_{\tau\tau}$ at the spacetime point of the particle, via
\begin{equation*}
g_{\tau\tau}(\tau,\rho)=-\frac{\|v_{\mathrm{Fermi}}\|^2}{\|v_{\mathrm{kin}}\|^2},
\end{equation*}
provided the denominator is not zero. Similarly, a measurement of the astrometric and spectroscopic speeds of a receding test particle  relative to the comoving observer  determine the metric tensor element $\tilde{g}_{\tau\tau}$ in optical coordinates at the spacetime point of the particle via
\begin{equation*}
\tilde{g}_{\tau\tau}(\tau,\delta)=\frac{1- \| v_{\mathrm{ast}}\|}{1+\|v_{\mathrm{spec}}\|}\|v_{\mathrm{spec}}\|-\frac{1+ \| v_{\mathrm{ast}}\|}{1+\|v_{\mathrm{spec}}\|}.
\end{equation*}
\end{proposition}
\begin{remark}\label{experiment}
It follows from Proposition \ref{findmetric} that in principle, knowledge of either pair of the relative velocities for moving test particles at each spacetime point uniquely determines the geometry of the two dimensional spacetime via \eqref{fermipolar} and \eqref{opticalmetric}, and therefore the scale factor $a(t)$. Since the affine distance (i.e the optical coordinate $\delta $) can be measured by parallax, and the frequency ratio can be found by spectroscopic measurements, the astrometric and spectroscopic relative velocities can be determined solely by physical measurements, and so, they could confirm or contradict assumptions about the value of $a(t)$ for the actual universe.
\end{remark}

Proposition \ref{findmetric} along with results of the preceding section allow us to calculate $g_{\tau\tau}(\tau,\rho)$ in terms of the curvature-normalized coordinates.  For example, equations \eqref{eqt1Galpha}, \eqref{powerVkinetic} and Proposition \ref{powerVFermi} immediately give the following result:

\begin{corollary} \label{remgtautau} For a Robertson-Walker spacetime with scale factor $a(t)=t^{\alpha}$, with $0<\alpha<1$, the metric tensor element $g_{\tau\tau}$ is given by
\begin{equation*}
g_{\tau\tau}(\tau,\rho)=-\left(G_{\alpha }^{1-\alpha }(v)+\frac{1-\alpha }{\alpha }\sqrt{1-G_{\alpha }^{2\alpha }(v) }\, v\right)^{2},
\end{equation*}
where $v=v(\tau,\chi)=\alpha\chi /\tau^{1-\alpha}$, and $\chi$ is the unique coordinate such that $(G_{\alpha }(v) \tau,\chi)$ in curvature-normalized coordinates represents the same event as $(\tau,\rho)$ in Fermi coordinates.
\end{corollary}

As an illustration of Corollary \ref{remgtautau}, for the radiation-dominated universe, i.e., when $\alpha = 1/2$, we immediately have from Example \ref{example2} that $g_{\tau\tau}(\tau,\rho)= -(\cos v +v \sin v)^{2}$,
and it may be verified using the coordinate transformation formulas in \cite{Klein11} that this expression is the same as \eqref{radgtt}.

\subsection{Retarded comparison for the spectroscopic relative velocity}
\label{secrel1}

The redshift of a radially moving test particle relative to the central observer is determined by the frequency ratio \eqref{doppleroptic}. In Fermi coordinates, the 4-momentum tangent vector field for the photon is defined on the light ray $\lambda $ and it is given by
\begin{equation}\label{photon}
\textrm{P}=\dot{\tau}\frac{\partial}{\partial\tau}+ \dot{\rho}\frac{\partial}{\partial\rho}=\frac{\mathcal{P}}{\sqrt{-g_{\tau\tau}}}\frac{\partial}{\partial\tau} -\mathcal{P}\frac{\partial}{\partial\rho},
\end{equation}
where the overdot represents differentiation with respect to the affine parameter of $\lambda $, and, analogous to $\mathcal{E}$ in \eqref{4vel},  $\mathcal{P}:=\sqrt{-g_{\tau\tau}}\,\dot{\tau}$ is the energy of the photon as measured by a stationary observer (in Fermi coordinates), i.e. an observer with 4-velocity $\frac{1}{\sqrt{-g_{\tau\tau}}}\frac{\partial}{\partial\tau}$.

Combining \eqref{doppleroptic}, \eqref{photon} and \eqref{4vel}, gives
\begin{equation}\label{shift}
\frac{\nu '}{\nu }= \frac{\mathcal{P}(q_{\ell })}{\mathcal{P}(p)}\left( \mathcal{E}^*\pm \sqrt{\mathcal{E}^{*2}-1}\right) =\frac{\nu'_0}{\nu}\left( \mathcal{E}^*\pm \sqrt{\mathcal{E}^{*2}-1}\right) ,
\end{equation}
where the $+$ or $-$ sign is taken for a receding or approaching (in Fermi coordinates) test particle respectively, the energy $\mathcal{E}^*$ is calculated for the test particle located at $q_{\ell }=(\tau ^*,\rho ^*)$, and $\nu'_0/\nu$ is the frequency ratio of a photon emitted from a stationary observer (whose 4-velocity is given by \eqref{u0}) from the event $q_{\ell }$. Using \eqref{vkinE}, \eqref{shift} may be expressed as
\begin{equation}\label{shift2}
\frac{\nu '}{\nu }= \frac{\nu '_0}{\nu }\mathcal{E}^*\left(1\pm\|v_{\mathrm{kin}}^*\|\right) ,
\end{equation}
where $v_{\mathrm{kin}}^*$ is the kinematic relative velocity of $u'_{\ell }$ measured from the event $p^*$ of $\beta $ with Fermi coordinates $(\tau^*,0)$ (see \eqref{vkinfermi}). The term $\nu '_0/\nu $ may be calculated by considering a  test particle comoving with the Hubble flow with fixed curvature-normalized coordinate $\chi$, for which the left side of \eqref{shift2} is known. In that case, the left side of \eqref{shift2} is given by \eqref{zshift}, i.e.,
\begin{equation}\label{doppler2}
\frac{\nu '}{\nu }=\frac{a(\tau)}{a(t_{\ell })},
\end{equation}
where $t_{\ell }=t_{\ell }(\tau ,\chi)$ is determined implicitly by \eqref{eqt1b}, and the coordinates $(t_{\ell },\chi)$ and $(\tau^{*},\rho^{*})$ describe the same event $q_{\ell }$.  The 4-velocity of the comoving test particle is given by $\left.\frac{\partial}{\partial t}\right|_{q_{\ell }}=\frac{\partial \tau}{\partial t_{\ell }}(t_{\ell }, \chi)\left. \frac{\partial}{\partial \tau}\right| _{q_{\ell }}+\frac{\partial\rho}{\partial t_{\ell }}(t_{\ell }, \chi)\left. \frac{\partial}{\partial\rho}\right| _{q_{\ell }}$.
Thus,
\begin{equation}\label{numerics}
\mathcal{E}^*=-g\left(\frac{1}{\sqrt{-g_{\tau\tau}(\tau^*,\rho^*)}}\left. \frac{\partial}{\partial\tau}\right| _{q_{\ell }},\left.\frac{\partial}{\partial t}\right|_{q_{\ell }}\right)=\sqrt{-g_{\tau\tau}(\tau^*,\rho^*)}\,\frac{\partial\tau}{\partial t_{\ell }}(t_{\ell },\chi).
\end{equation}
The partial derivative on the right side of \eqref{numerics} may be calculated as follows. Since $\chi $ is fixed, we may write $t_{\ell }=t_{\ell }(\tau)$. Starting with the coordinate transformation formula, $\tau(t_{\ell }(\tau), \chi)=\tau^{*}$, we have by the chain rule,
\begin{equation}
\label{partialderiv}
\frac{d\tau^{*}}{d\tau}=\frac{\partial \tau}{\partial t_{\ell }}(t_{\ell }, \chi)\frac{dt}{d\tau}
\quad \Longrightarrow \quad
\frac{\partial \tau}{\partial t_{\ell }}(t_{\ell }, \chi)=\frac{\,d\tau^{*}}{d\tau}\frac{d\tau}{\,dt_{\ell }}.
\end{equation}
Next we find $d\tau^{*}/d\tau$. Since the vector field $\textrm{P}$ is tangent to $\lambda $, using \eqref{photon} we find
\begin{equation}\label{t*}
\int _{\tau^*}^{\tau}\,\textrm{d}\bar{\tau } =\int_{\rho^{*}}^{0}\frac{-1}{\sqrt{-g_{\tau\tau}(\tau(\bar{\rho }),\bar{\rho })}}\,\textrm{d}\bar{\rho }\quad \Longrightarrow \quad \tau =\tau^{*}+\int^{\rho^*(\tau^{*})}_{0}\frac{1}{\sqrt{-g_{\tau\tau}(\tau(\bar{\rho }),\bar{\rho })}}\,\textrm{d}\bar{\rho },
\end{equation}
which determines $\tau$ as a function of $\tau^{*}$ with
\begin{equation}\label{dtdtau*}
\frac{d\tau\,}{d\tau^{*}}=1+ \frac{1}{\sqrt{-g_{\tau\tau}(\tau^{*},\rho^{*})}}\frac{d\rho^*}{d\tau^{*}}=1+\|v_{\mathrm{kin}}^*\|>0.
\end{equation}
It follows that \eqref{t*} determines the inverse function, $\tau^{*}(\tau)$ as well. On the other hand, differentiating \eqref{eqt1b} with respect to $\tau $ and taking into account that $\chi $ is constant, we obtain
\begin{equation}
\label{t1dot}
\frac{dt_{\ell }}{d\tau}=\frac{a(t_{\ell })}{a(\tau)}.
\end{equation}
Using \eqref{dtdtau*} and \eqref{t1dot} to calculate the right side of \eqref{partialderiv}, we have $\frac{\partial \tau}{\partial t_{\ell }}(t_{\ell }, \chi)= \frac{a(\tau)}{a(t_{\ell })\left(1+\|v_{\mathrm{kin}}^*\|\right)}$.
Thus, from \eqref{numerics}, we have for the comoving test particle,
\begin{equation}\label{comovingenergy}
\mathcal{E}^*=\frac{a(\tau )\sqrt{-g_{\tau\tau}(\tau^*,\rho^*)}}{a(t_{\ell })\left(1+\|v_{\mathrm{kin}}^*\|\right)}.
\end{equation}
Combining \eqref{comovingenergy} and \eqref{doppler2} with \eqref{shift2}, yields the frequency ratio of a photon emitted by a stationary observer (in Fermi coordinates) at $q_{\ell }=(\tau^*,\rho^*)$ and received by the Fermi observer at $p=(\tau,0)$,
\begin{equation}\label{staticshift}
\frac{\nu'_0}{\nu}=\frac{1}{\sqrt{-g_{\tau\tau}(\tau^*,\rho^*)}}.
\end{equation}
Now combining \eqref{staticshift} with \eqref{shift2} gives the following expression for a radially moving particle (not necessarily comoving with the Hubble flow):
\begin{equation}\label{shift3}
\frac{\nu'}{\nu}= \frac{\mathcal{E}^*\left(1\pm\|v_{\mathrm{kin}}^*\|\right)}{\sqrt{-g_{\tau\tau}(\tau^*,\rho^*)}}.
\end{equation}
The corresponding spectroscopic relative velocity may be computed and expressed in terms of the kinematic relative velocity for a radially moving particle, directly from \cite[Equation (12)]{Bolos07} and \eqref{shift2} above, as
\begin{equation} \label{vspec}
v_{\mathrm{spec}}=\frac{(\nu'/\nu)^{2}-1}{(\nu'/\nu)^{2}+1}\mathcal{S}_p.
\end{equation}
Specializing, for the sake of simplicity, to the case of a radially receding test particle \eqref{vspec} may be rewritten as
\begin{equation}\label{compare2}
\frac{\nu'}{\nu}=\sqrt{\frac{1+\|v_{\mathrm{spec}}\|}{1-\|v_{\mathrm{spec}}\|}}.
\end{equation}
Now, combining \eqref{compare2} with \eqref{shift3} yields an expression for $\|v_{\mathrm{kin}}^*\|$ in terms of $\|v_{\mathrm{spec}}\|$,
\begin{equation}\label{compare3}
\|v_{\mathrm{kin}}^*\|=\frac{\sqrt{-g_{\tau\tau}(\tau^*,\rho^*)}}{\mathcal{E}^*}\sqrt{\frac{1+\|v_{\mathrm{spec}}\|}{1-\|v_{\mathrm{spec}}\|}}-1.
\end{equation}

\subsection{Retarded comparison for the astrometric relative velocity}
\label{secrel2}

The modulus of the astrometric relative velocity of a radially moving test particle relative to $\beta $ is $|d\delta /d\tau| $, where $\delta $ is the affine distance from $p$ to $q_{\ell }$ (see Figure \ref{diagram}).
To compute this we use the fact that $\lambda(\delta ) = q_{\ell }$, where $\lambda (\xi)$ is an affinely parameterized null geodesic from $\lambda (0)=p$ to $q_{\ell }$, with $\left. \frac{d\lambda}{d\xi}\right| _0\equiv\textrm{P}_p=-\left. \frac{\partial}{\partial\tau}\right| _p+ \left. \frac{\partial}{\partial\rho}\right| _p$ (see \cite[Propositions 6 and 7]{Bolos07}) and whose tangent vector field $\textrm{P}$ is given by \eqref{photon}. Therefore, $\rho^{*}= \int^{\delta }_{0}\dot{\rho}\,\textrm{d}\xi$, where the overdot represents differentiation with respect to  $\xi $, and thus
\begin{equation*}
\frac{d\rho^{*}}{\!\!d \delta}=\dot{\rho}(\delta)=-\sqrt{-g_{\tau\tau}(\tau^{*},\rho^{*})}\,\,\dot{\tau}(\delta )=g(u_0,\textrm{P}_{q_{\ell }})=-\mathcal{P}(q_{\ell }),
\end{equation*}
where $u_0$ is the stationary observer at $q_{\ell }$. Note that the previous expression is positive since the photon is backward traveling.  In order to calculate $\mathcal{P}(q_{\ell })$, consider a forward moving photon traveling from $q_{\ell }=(\tau^{*}, \rho^{*})$ to $p=(\tau ,0)$ following the future-pointing null geodesic $\tilde{\lambda}(\xi)\equiv\lambda(\delta -\xi)$. Then \eqref{doppleroptic} and \eqref{staticshift} give $-\mathcal{P}(q_{\ell })=\frac{\nu '_0}{\nu }=\frac{1}{\sqrt{-g_{\tau\tau}(\tau^*,\rho^*)}}$, and thus,
\begin{equation}\label{affineFermi}
\frac{\!\!d\delta }{d \rho^{*}}=\sqrt{-g_{\tau\tau}(\tau^*,\rho^*)}.
\end{equation}
Using the chain rule together with \eqref{dtdtau*}, \eqref{affineFermi} and the $^*$-version of \eqref{fermidefinition2}, it follows that the modulus of the astrometric relative velocity of $u'_{\ell }$ observed by $u$ is given by

\begin{equation*}\label{ast'}
\| v_{\mathrm{ast}}\| = \left| \frac{d\delta }{d\tau }\right|
 = \frac{\,\,d\delta }{d\rho ^*}\left| \frac{d\rho ^*}{d\tau ^*}\right| \frac{d\tau ^*}{d\tau }=\left|\frac{d\rho^{*}}{d\tau^{*}}\right|\frac{\sqrt{-g_{\tau\tau}(\tau^*,\rho^*)}}{1+ \frac{1}{\sqrt{-g_{\tau\tau}(\tau^{*},\rho^{*})}}\frac{d\rho^{*}}{d\tau^{*}}},
\end{equation*}
which may be expressed in terms of the Fermi or kinematic relative velocity of $u'_{\ell }$ measured from the event $p^*$ of $\beta $ with Fermi coordinates $(\tau^*,0)$ as,
\begin{equation}
\label{ast}
\| v_{\mathrm{ast}}\|  = \frac{\sqrt{-g_{\tau \tau }(\tau ^*,\rho ^*)}\|v_{\mathrm{Fermi}}^*\|}{1\pm \frac{\| v_{\mathrm{Fermi}}^*\|}{\sqrt{-g_{\tau\tau}(\tau ^*,\rho ^*)}}}=\frac{-g_{\tau\tau}(\tau^{*},\rho^{*})\,\|v_{\mathrm{kin}}^*\|}{1\pm\|v_{\mathrm{kin}}^*\|}\,,
\end{equation}
where the positive sign is used for a radially receding particle, and the negative sign for a radially approaching particle, and where for the second equality we used \eqref{compare5}.

Finally, we can find relationships between the astrometric and spectroscopic relative velocities. For ease of exposition, we specialize to the case of a receding test particle. Rearranging \eqref{ast}, we have
\begin{equation*}
\|v_{\mathrm{kin}}^*\|=\frac{\|v_{\mathrm{ast}}\|}{-g_{\tau\tau}(\tau^{*},\rho^{*})-\|v_{\mathrm{ast}}\|},
\end{equation*}
and then from \eqref{compare3}, we find
\begin{equation*}
\label{vastend1}
\|v_{\mathrm{ast}}\|=-g_{\tau\tau}(\tau^*,\rho^*)\left( 1-\frac{\mathcal{E}^*}{\sqrt{-g_{\tau\tau}(\tau^*,\rho^*)}}
\sqrt{\frac{1-\|v_{\mathrm{spec}}\|}{1+\|v_{\mathrm{spec}}\|}}\right) .
\end{equation*}
This may be compared with the following expression in optical coordinates, which follows from \eqref{vastoptical} and \eqref{vspecinoptical} (see also \eqref{astro/spec}),
\begin{equation*}
\label{vastend2}
\|v_{\mathrm{ast}}\|=-\frac{1}{2}\left(\tilde{g}_{\tau\tau}(\tau,\delta)+
\frac{1-\|v_{\mathrm{spec}}\|}{1+\|v_{\mathrm{spec}}\|}\right) .
\end{equation*}

\section{Concluding remarks}
\label{conclusion}

We have found general expressions for the Fermi, kinematic, astrometric, and spectroscopic  velocities of test particles experiencing radial motion relative to an observer  comoving with the Hubble flow (called the central observer) in any expanding Robertson-Walker cosmology.  Specific numerical calculations and formulas were given for the de Sitter universe and cosmologies for which the scale factor, $a(t)=t^{\alpha}$ with $0<\alpha\leq1$, including, in particular, the radiation-dominated and matter-dominated universes.

In Proposition \ref{findmetric}, we showed how pairs of these relative velocities, for arbitrary test particles, determine the leading metric tensor coefficients in Fermi or optical coordinates, and in Remark \ref{experiment}, we observed that the astrometric and spectroscopic relative velocities could be used, at least in principle, to measure the scale factor $a(t)$ of the actual universe.  An analog of Hubble's law for the astrometric relative velocity of comoving particles was given in Proposition \ref{hubble}.

Of the four relative velocities, only the Fermi relative velocity of a radially receding test particle can exceed the speed of light, and this is possible at a spacetime point $(\tau, \rho)$, in Fermi coordinates, if and only if $-g_{\tau\tau}(\tau, \rho)>1$. Examples of superluminal Fermi velocities were given in Section \ref{examples}.

Under general conditions, the Hubble velocity of comoving test particles also become superluminal at large values of the radial parameter, $\chi$, and this is taken as a criterion for the expansion of space in cosmological models, and for the actual universe. By way of comparison, the Fermi relative velocity has both advantages and disadvantages to the Hubble velocity.  For comoving particles, both velocities measure the rate of change of proper distance away from the observer with respect to the proper time of the observer.  But for the Fermi velocity, the proper distance is measured along spacelike geodesics, while for the Hubble velocity the proper distance is measured along non geodesic paths.  In this respect the Fermi velocity is more natural and more closely tied to the observer's natural frame of reference, i.e., to Fermi coordinates in which locally the metric is Minkowskian to first order in the coordinates.  In addition, the notion of Fermi relative velocity (as well as the other three relative velocities considered herein) is geometric and applies to any spacetime, while the Hubble velocity is specific to Robertson-Walker cosmologies.

In its favor, the Hubble velocity is defined at all spacetime points, whereas the Fermi velocity makes sense only on the Fermi chart of the central observer.  For the case of non inflationary cosmologies, such as when the scale factor follows a power law, $a(t)=t^{\alpha}$, the Fermi chart is global \cite{Klein11}, and the Fermi velocity does not suffer this disadvantage.

However, for inflationary cosmologies such as the de Sitter universe, the Fermi chart is valid only up to the cosmological horizon \cite{CM,KC3,Klein11}.  For this particular example, the behavior of the Fermi and Hubble velocities of comoving particles is qualitatively different. Fermi relative velocities do not reach half the speed of light, while Hubble velocities become superluminal, as shown in Figure \ref{vdesitter_all}.  This is because the notions of simultaneity are different for the corresponding coordinate systems, and the two velocities are ``measuring'' the expansion of different hyperspaces.  It would be interesting and possibly useful to have a rigorous and universal mathematical criterion for expansion of space.

\end{document}